# Nonperturbative Studies of Quantum Gravity

J. Riedler
Institut für Kernphysik
Technische Universität Wien
A-1040 Vienna, Austria

## Abstract

One of several possibilities to construct a quantum theory of gravity is employing the Feynman path integral. This approach is plagued by some problems: the integration measure is not uniquely defined, the Einstein-Hilbert action unbounded, and perturbation theory nonrenormalizable. To make the path integral tractable one can approximate the continuous geometry of spacetime by a simplicial complex. The edge lengths of this lattice are considered as the dynamical degrees of freedom and Regge calculus is applied. In this work, numerical simulations using the Regge-Einstein action and a "compact" action show the occurence of a phase transition. The strength of this transition, separating a well-defined phase with finite expectation values from an ill-defined phase, is weaker for the compact action, which might be important for the continuum limit. To analyze the interaction mechanism of this formulation of quantum gravity, correlation functions between geometrical quantities like edge lengths, volume elements, and local curvatures have been computed. Our results for the two-point functions seem to prefer exchange particles with an effective mass. To ease treatment of quantum gravity a new approach is proposed consisting in a transformation of the path integral to the partition function of a spin system. This facilitates analytical and numerical calculations considerably. First results for the phase structure in two as in four dimensions are presented and indicate promising similarities to the original Regge theory.

# Contents

# 1 Technical and Conceptual Problems of Quantum Gravity

At present, with quantum chromodynamics for the strong interaction, the Glashow-Salam-Weinberg model for the electromagnetic and weak interactions, and Einstein's theory of general relativity for gravitation one may feel to have arrived at a closed system of physical laws. However, problems arise when trying to describe these theories in a unified framework. Although particle physics is successfully described by quantum field theories, the laws of gravity determining the large scale structures of the universe seem to be different. Based on the assumption that quantum theory is universally valid it would appear strange to have a drastically different framework for gravitation. Notwithstanding many decades of intense work, there is still no complete quantum theory of gravity with the properties thought to be essential for consistency, such as unitarity, renormalizability (or some related property, like finiteness or asymptotic safety) and Lorentz invariance in a local inertial frame. The difficulties are compounded by the total lack of any empirical data that is manifestly relevant to the problem and a serious obstacle for experiments in this field is the extreme smallness of the Planck length

$$L_P = \sqrt{\frac{\hbar G}{c^3}} \approx 10^{-35} \text{m} .\qquad(1)$$

There are three especially important issues confronting every approach to quantum gravity [1, 2, 3].

## 1.1 Background structure

The mathematical model of spacetime used in classical general relativity is a differentiable manifold equipped with a Lorentzian metric. Now one can ask for the underlying substructure of this picture.

The bottom level is a set $M$ whose elements are to be identified with spacetime points or events. This set is formless with its only general mathematical property being the cardinal number. In particular, there are no relations between the elements of $M$ and no special way of labeling any such element. The next step is to impose a topology on $M$ so that each point acquires a family of neighborhoods. It now becomes possible to talk about relationships between points, albeit in a rather nonphysical way. This defect is overcome by adding the key ingredient of all standard views of spacetime: the topology of $M$ must be compatible with that of a differentiable manifold. A point can then be labeled uniquely in $M$ (at least locally) by giving the values of $n$ real numbers, with $n$ the dimension of the manifold. Such a coordinate system also provides a more specific way of describing relationships between points of $M$, though not intrinsically in so far as these depend on which coordinate systems are chosen to cover $M$. In the final step a Lorentzian metric $g$ is placed on $M$, thereby introducing the ideas of the length of a path joining two spacetime points, parallel transport with respect to a Riemannian connection, causal relations between pairs of points etc.

However, since one wishes to assert that some sort of quantum spacetime structure is meaningful, the key question for any particular approach to quantum gravity is how much of the hierarchy described above must be kept fixed. For example, in most of the former approaches to quantizing the gravitational field, the set of spacetime points, topology and differential structure are all fixed, and only the Lorentzian metric $g$ is subject to quantum fluctuations. If the Lorentzian metric $g$ becomes quantized then the light cone associated with any spacetime point is no longer fixed and it is not meaningful to impose microcausal relations. This destroys one of the bedrocks of conventional quantum field theory and is probably the greatest reason why spacetime approaches to quantum gravity have not got as far as might have been hoped.

## 1.2 Spacetime diffeomorphism group

The group $\text{Diff}(M)$ of spacetime diffeomorphisms plays a key role in the classical theory of general relativity and so the question of its status in quantum gravity is of considerable interest. The action of $\text{Diff}(M)$ on $M$ affects the space $\mathcal{F}$ of spacetime fields, and the only thing that has immediate physical meaning is the quotient space $\mathcal{F}/\text{Diff}(M)$ of orbits, i.e. two field configurations are regarded as physically equivalent if they are connected by a $\text{Diff}(M)$ transformation. Technically, this is analogous to the situation in electromagnetism where a vector potential $A_\mu$ is equivalent to $A_\mu + \partial_\mu f$ for all functions $f$. However, there is an important difference between electrodynamics and general relativity. Electromagnetic gauge transformations occur at a fixed spacetime point $x$, and the physical configurations can be identified with the values of the electromagnetic field $F_{\mu\nu}(x)$, which depends locally on points of $M$. On the other hand, $\text{Diff}(M)$ maps one spacetime point into another, and therefore the obvious way of constructing a $\text{Diff}(M)$-invariant object is to take some scalar function of spacetime fields and integrate it over the whole of $M$, which gives something that is very nonlocal. The idea that physical observables are naturally nonlocal is an important ingredient in many approaches to quantum gravity.

## 1.3 Time

One of the major issues in quantum gravity is the so-called problem of time. This arises from the very different roles played by the concept of time in quantum theory and in general relativity. In standard quantum theory time is not a physical observable in the normal sense since it is not represented by an operator. Rather, it is treated as a parameter which, as in classical physics, is used to mark the evolution of the system. In particular, it provides the parameter $t$ in the time-dependent Schrödinger equation

$$i\hbar \frac{d\psi}{dt} = \hat{H}\psi \ . \qquad (2)$$

In addition, the scalar product of states and complete sets of commuting observables are defined at a single time, measurements are carried out at a single time, and states specify the probability of the results of such measurements.

In a Diff($M$)-invariant theory like classical general relativity the role of time is very different. If $M$ is equipped with a Lorentzian metric $g$, and if its topology is appropriate, it can be foliated in many ways as a 1-parameter family of spacelike surfaces, and each such parameter might be regarded as a possible definition of time. Such a definition of time is rather unphysical since it provides no hint as to how it might be measured or registered. Moreover, the possibility of defining time in this way is closely linked to a fixed choice of the metric $g$. It becomes untenable if $g$ is subject to some type of quantum fluctuation.

One might try to construct a theory of quantum gravity by using functional integrals analogous to those employed in normal quantum field theory to produce the vacuum expectation values of the time-ordered product of fields. However, it is difficult to see what a time-ordered product could mean in the absence of any background metric to provide a preferred notion of timelike and spacelike.

Attempts to construct a quantum theory of gravity can be divided into three broad categories briefly described in the following Sections.

## 1.4 Quantizing general relativity

The idea is to start with the classical theory of general relativity represented by the Einstein-Hilbert action

$$I[g] = \frac{1}{16\pi G} \int d^4x \sqrt{-g(x)} R(g) \tag{3}$$

($G$ labels Newton's constant, $g$ the determinant of the metric tensor, and $R$ the curvature scalar) and then to apply some type of quantization algorithm. This is intended to be analogous to the way in which the classical theory of an electron bound by the Coulomb potential is quantized by replacing certain classical observables with self-adjoint operators on a Hilbert space. Of course, this is essentially also the approach used in developing important elementary-particle physics ideas like the Glashow-Salam-Weinberg electro-weak theory and the quantum chromodynamics description of the strong nuclear force.

In early particle-physics based approaches to quantum gravity the starting point is to fix the background topology and differential structure of spacetime $M$ to be that of Minkowski space, and then to write the Lorentzian metric $g$ on $M$ as

$$g_{ab}(x) = \eta_{ab} + h_{ab}(x) \,, \tag{4}$$

where $h$ measures the departure of $g$ from flat spacetime $\eta$. The background metric $\eta$ provides a fixed causal structure with the usual family of Lorentzian inertial frames. Thus, at this level, there is no problem of time. The causal structure also allows a notion of microcausality, thereby permitting a conventional type of relativistic quantum field theory to be applied to the field $h_{ab}$. The quanta of this field are massless spin-2 particles, the gravitons. A typical task would then be to compute perturbative scattering-matrix elements for these gravitons. The action of Diff($M$) is usually studied infinitesimally and is reflected in the quantum theory via a set of Ward identities that must be satisfied by the $n$-point functions of the theory. Unfortunately, ultraviolet divergences are sufficiently violent to render the theory perturbatively nonrenormalizable [4].

One of many responses was to enlarge the classical theory of general relativity with carefully chosen matter fields with the hope that the ultraviolet divergences would cancel, leaving a theory that is perturbatively well-behaved. The concept is based on the extension of the group of global Poincaré transformations with anticommuting spinor generators $Q$. Hence, a supersymmetry between bosons and fermions is achieved, which allows to transform bosonic to fermionic states and vice versa

$$Q|B\rangle \sim |F\rangle \ , \quad Q|F\rangle \sim |B\rangle \ . \tag{5}$$

Thus, each particle obtaines a corresponding SUSY partner. Indeed, divergent bosonic (fermionic) loop contributions can compensate those of fermions (bosons). A generalization to local supersymmetric transformations leads to a gauge theory of supersymmetry, called supergravity [5]. The Lagrangian of simple supergravity reads

$$\mathcal{L} = -\frac{1}{16\pi G} e R(e,\omega) - \frac{1}{2}\epsilon^{klmn}\bar\psi_k \gamma_5 \gamma_l D_m \psi_n \ , \tag{6}$$

with $e$ the determinant of the vielbein, $\omega$ the spin connection, and $\gamma_i$ the Dirac matrices. The gravitino field $\psi_m$ is the SUSY partner of the graviton and is associated to the generator $Q$. Early expectations were following successful low-order results but it is now generally accepted that if higher-loop calculations could be performed intractable divergences would appear once more.

The fact that a theory is perturbatively nonrenormalizable does not necessarily mean that it has an intrinsic problem or that it is bad in any way. It merely says that perturbation theory does not apply to the problem in question. It is therefore not surprising that a great deal of developments focus on nonperturbative quantization in the context of the canonical theory.

The first step in canonically quantizing general relativity consists in foliating spacetime into a family of space-like 3-dimensional hypersurfaces, i.e. one decomposes spacetime into space and time. The metric on these hypersurfaces, $h_{ab}(\mathbf{x})$, will play the role of the canonical variable. All quantities are then decomposed into variables which exist on such a hypersurface and variables which point in the fourth, timelike dimension. It then turns out that the canonical momentum, $\Pi^{ab}(\mathbf{x})$, is given by the extrinsic curvature of a 3-dimensional hypersurface, i.e. the quantity which describes the embedding of space into spacetime. As a consequence of invariance under arbitrary coordinate transformations one finds that there exist four constraints, of which the so-called Hamiltonian constraint $\mathcal{H}$ is directly connected to the invariance of the theory under reparametrizations of time. Its explicit form is

$$\mathcal{H} \equiv \frac{16\pi G}{c^2} G_{abcd}\Pi^{ab}\Pi^{cd} - \frac{c^4}{16\pi G}\sqrt{h}R + \mathcal{H}_m = 0 \ , \tag{7}$$

where R is the curvature scalar on 3-space and $\mathcal{H}_m$ is the Hamiltonian density for nongravitational fields. The coefficients $G_{abcd}$ depend explicitly on the metric and play the role of a metric in the space of all metrics. Quantization now proceeds, by elevating the metric and its momentum to the status of operators and imposing the commutation relations

$$[h_{ab}(\mathbf{x}), \Pi^{cd}(\mathbf{y})] = i\hbar \delta_a^{\ c} \delta_b^{\ d} \delta(\mathbf{x} - \mathbf{y}) \ . \tag{8}$$

One specific realization of (8) is provided by the substitution

$$\Pi_{ab} \to \frac{\hbar}{i} \frac{\delta}{\delta h_{ab}} \ . \tag{9}$$

The classical constraint (7) is then formally implemented in the quantum theory by inserting (9) into (7) and applying it on wave functionals $\Psi$ which depend on the 3-metric and on nongravitational fields denoted by $\phi$, i.e.

$$\mathcal{H}\Psi[h_{ab}(\mathbf{x}), \phi(\mathbf{x})] = \left( -\frac{16\pi\hbar^2 G}{c^2} G_{abcd} \frac{\delta^2}{\delta h_{ab} h_{cd}} - \frac{c^4}{16\pi G} \sqrt{h} R + \mathcal{H}_m \right) \Psi = 0 \ . \tag{10}$$

This is the well-known Wheeler-DeWitt equation [6], the gravitational analog of the Schrödinger equation. The quantization of the remaining constraints leads to the condition that this wave functional does not change under coordinate transformations of the 3-metric, but is a function of the geometry only. The configuration space is thus the space of all 3-geometries and is called superspace. Several techniques have been suggested for finding solutions to the Wheeler-DeWitt equation. An interesting development has been the discovery of appropriate variables which enables to find exact solutions of (7) in the absence of matter. This became possible since the complicated potential term of (7) disappears when it is rewritten in terms of these Ashtekar variables [7], which have a strong similarity to Yang-Mills loop variables. The solutions can be classified in terms of loops and knots and exhibit an interesting structure of space, of which the most important may be the existence of a minimal length [8].

Another nonperturbative scheme is provided by the functional integral approach. Here only a brief overview is given, but it will be outlined in more detail in the next Chapter. In ordinary quantum mechanics a solution to the time-dependent Schrödinger equation (2) can be generated in the form

$$\psi(x, t) = \int D[x(s)] e^{\frac{i}{\hbar} \int_{t_0}^{t} L(s) ds} \ , \tag{11}$$

where the integral is over all paths that end at the point $x$ at time $t$. The solution thus obtained depends on the value $x(t_0)$ of the path at the initial time $t_0$.

The gravitational analog of (11) plays an important role in the Hartle-Hawking approach to quantum gravity [9]. There, the wave functional is expressed as a formal path integral

$$\Psi[h] = \int_{\partial M} D[g] e^{-I_E[g]} \ , \tag{12}$$

where the sum is over Euclidean geometries $g$ on the 4-manifold $M$. The boundary $\partial M = \{\Sigma_i, \Sigma_f\}$ consists of 3-dimensional hypersurfaces with induced metrics $h$. $I_E[g]$ is the Euclidean form of (3) and contains an additional boundary term. Note, that there is no explicit notion of time in (12) and the proper time between the surfaces depends on the 4-geometries in the sum.

When applied to cosmology, $\Psi$ is the wave function of the universe. The main motivation for this claim comes from the nonseparability of quantum theory, i.e. from the fact that one cannot in general assign a wave function to a given system because it is not isolated but coupled to its natural environment, which is again coupled to another environment, and so forth. The extrapolation of this quantum entanglement leads inevitably to the concept of a wave function for the universe.

It can be shown that (12) satisfies the Wheeler-DeWitt equation and again the solution depends on the boundaries. In the Hartle-Hawking *no boundary ansatz* it is assumed that one has to perform a sum over compact manifolds with one boundary only, which is given by the considered universe. The lack of a second boundary (at a small size of the universe) saves one from the need to find appropriate boundary conditions there: the universe is created *ex nihilio*.

However, the quantum-gravity path-integral is also plagued with a lot of problems concerning the integration over 4-geometries, the Euclidean gravitational action and the measure. In particular, the integration should include a summation over manifolds of different topologies, but topologies are not classifiable in 4 dimensions. Due to conformal fluctuations, the gravitational action is not bounded from below, which results at a first view in a divergent path integral. A unique definition of the measure does not exist because different physical motivations lead to different measures.

Nevertheless, approximate calculations of functional integrals have been explored to some extent. These include minisuperspace models obtained by freezing all but a finite number of modes of the 4-geometry, e.g. lattice gravity approaches like Regge calculus [10].

The idea of approximating spacetime by a simplicial complex has been of interest in both classical and quantum gravity for a long time. The Regge calculus provides a direct route in approximating continuous 4-geometries by simplicial lattices [11]. Expectation values can be written as

$$\langle \mathcal{O} \rangle = \sum \nu(K) \int d\mu[l] \mathcal{O}(l) e^{-I_R(l)} , \tag{13}$$

where $\nu(K)$ is the measure weight-factor of the simplicial complex $K$ with $N_0$ vertices, $\mu[l]$ is the measure and $\mathcal{O}(l)$ is an observable written as a function of the edge lengths $l$ of the simplices. The Regge analog $I_R$ of the Einstein-Hilbert action (3) is usually given as a function of the area $A_t$ of the 2-simplices $t$ and of the deficit angles $\delta_t$ of the 4-simplices containing these faces

$$I_R(l) = \frac{1}{8\pi G} \sum_t A_t \delta_t . \tag{14}$$

As the notation suggests it can be rewritten as a complicated function of the edge lengths alone. It is hoped that the classical continuum limit will be achieved as $N_0 \to \infty$ and $l \to 0$. The other possibility would be a finite length scenario with the attractive features of avoiding difficulties with locality at the Planck length and of providing a natural ultraviolet cut-off making renormalization of infinities unnecessary [12]. But still, many of the intriguing questions and problems arising in the conventional (continuum) functional integral approach are also present in the Regge calculus.

## 1.5 "General-relativize" quantum theory

The main idea is to begin with some prior concept of quantum theory and then to force it to be compatible with general relativity. The biggest programme of this type is due to Fredenhagen and Haag [13]. It can be shown that the basic tenets of the general theory of relativity, namely general covariance and strict locality can be incorporated into quantum theory. Using the assumption that the theory has a scaling limit, each quantum state defines a reduced theory in the tangent space of each point. This reduced theory turns out to be invariant under translations and even under $SLR(4)$. Then the macroscopic metric should evolve as a cooperative effect in finite size regions. However, many questions remain to be answered before a viable theory can be proposed.

## 1.6 General relativity as a low-energy limit

Another possibility is to remain with the idea of quantizing a classical system but start with something else than general relativity. The key step here is to find a system whose quantum theory is well-defined and which yields classical general relativity as a low-energy limit. A very sophisticated example for a scheme like this is the string theory, which abandons the concept of idealized point particles. The fundamental entities are 1-dimensional objects (strings) instead of local quantum fields and the graviton occurs as just one of an infinite number of particles associated with the quantized string [14].

In the bosonic case one has the Polyakov action

$$I[g, X] = \frac{1}{4\pi\alpha'} \int_W d^2\sigma \sqrt{q(\sigma)} q^{ij}(\sigma) \partial_i X^\mu(\sigma) \partial_j X^\nu(\sigma) g_{\mu\nu}(X) \ , \qquad (15)$$

where $q_{ij}$ is a metric on the 2-dimensional manifold $W$ (the world sheet) parametrized by $\sigma = (\sigma_1, \sigma_2)$, and $X : W \to M$ are the bosonic string fields which map $W$ into the spacetime manifold $M$ (the target space) with background metric $g_{\mu\nu}$. The Regge slope $\alpha'$ is related to the string tension $T = \frac{1}{2\pi\alpha'}$ and is assumed to be of the order of the Planck length.

A supersymmetric extension, the so-called heterotic string theory is formulated in a 10-dimensional world and obviously fails to reproduce the fact that we live in a 4-dimensional almost flat Minkowski spacetime. Thus 10-dimensional superstring theory has to be compactified: six coordinates are curled up describing a tiny compact space whose size is of the order of the Planck length, whereas the remaining physical four coordinates are kept uncompactified. Although it was recently discovered how to construct string theories using conformal field theory techniques without referring to any compactification scheme, there exists so far no compelling principle to determine the number of spacetime dimensions to be four. All dimensions below ten seem to be on an equal footing.

Superstring theory has the great advantage over the covariant approaches that the individual terms in the appropriate perturbation expansion *can* be finite and, furthermore, the particle content of theories of this type could well be such as to relate the fundamental forces in a unified way. The low-energy limit of these theories is a form of supergravity because, when the string tension tends to infinity ($\alpha' \to 0$) the strings degenerate into points. Nevertheless, standard spacetime ideas in the sense of general relativity do not play a very significant role. This is reflected by the graviton being only one of an infinite number of particles in the theory. Similarly, the spacetime diffeomorphism group appears only as part of a much bigger symmetry structure.

# 2 Sum-over-Histories Approach

In this work a nonperturbative treatment of quantum gravity using the path integral is persued. The essential point of this approach is Feynman's idea to represent the transition amplitude from an initial state with metric $h_1$ on a surface $\Sigma_1$ to a final state with $h_2$ on $\Sigma_2$ as a sum over all field configurations $[g]$ which take the given values on the surfaces $\Sigma_1$ and $\Sigma_2$ [3]

$$\langle h_2 | h_1 \rangle = \int_M D[g] e^{iI[g]} , \qquad (16)$$

cf. Figure 1. The integration extends over all 4-geometries with the measure $D[g]$, and $I[g]$ denotes the gravitational action, which in general relativity is usually chosen as

$$I = \frac{1}{16\pi G} \int_M d^4x \sqrt{-g}(R - 2\Lambda) . \qquad (17)$$

$R$ is the curvature scalar, $\Lambda$ the cosmological constant, and $g$ the determinant of the metric. $G$ is Newton's constant and units are such that $c = \hbar = 1$.

Strictly speaking the gravitational action contains second derivatives of the metric which have to be removed by integration by parts to give an action quadratic in first derivatives as is required by the path integral approach. Therefore, one gets a surface term

$$\frac{1}{8\pi G} \int_{\partial M} d^3x \sqrt{\pm h} K + C , \qquad (18)$$

where $K$ is the trace of the extrinsic curvature of the boundary $\partial M$ with the induced metric $h$. The plus or minus sign is chosen according to whether the boundary is spacelike or timelike. This term turns out to be important to fulfill the composition law for the amplitude to go from the initial to the final state, to be obtained by summing over all states on some intermediate surface $\Sigma_3$ with metric $h_3$ (see Figure 1)

$$\langle h_2 | h_1 \rangle = \sum_{h_3} \langle h_2 | h_3 \rangle \langle h_3 | h_1 \rangle . \qquad (19)$$

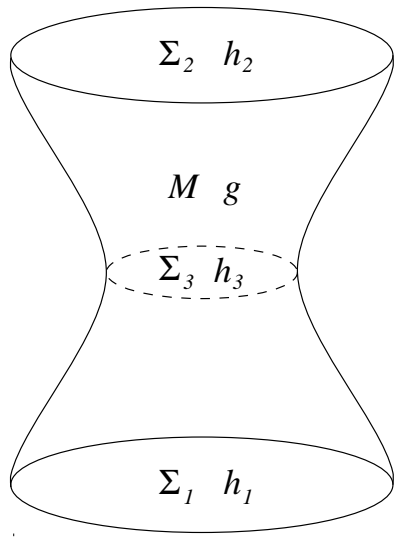

Figure 1: Manifold $M$ with metric $g$ and the boundary $\partial M$ consisting of $\Sigma_1$ and $\Sigma_2$ on which the metrics $h_1$ and $h_2$ are induced.

Under variations of the metric the action (17) is stationary if the Einstein equations in vacuum

$$R_{mn} - \frac{1}{2} g_{mn} R + \Lambda g_{mn} = 0 , \qquad (20)$$

with $R_{mn}$ the Ricci tensor, are satisfied.

## 2.1 Gravitational action

For real Lorentzian metrics $g$ the action $I[g, \phi]$ will be real and so the integrand in the path integral will oscillate. In ordinary quantum field theory in flat spacetime one can circumvent this difficulty by performing a Wick rotation into the complex plane. It seems reasonable to apply similar ideas to the gravitational field. The analog of the Wick rotation procedure is to perform the integration over metrics of positive definite signature. However, unlike the case of flat space quantum theory, this procedure is rather more than giving a mathematical meaning to the functional integral. The reason is that not all Lorentzian metrics have Euclidean sections, and vice versa. Therefore, if the integral is performed over all Euclideanized metrics then some Lorentzian metrics will have been omitted, whilst if the integration is taken over all Euclidean metrics then this will include some which do not admit Lorentzian sections, and are thus not physical. Hawking [3] has proposed a radical solution to this problem: the integral should be performed over all Euclidean metrics, and this can be regarded as being rather like a contour integral in the space of all complex metrics in which the contour has been rotated from the Lorentzian to the Euclidean section. Thus individual metrics in the functional integral need not necessarily have a direct physical interpretation and only $Z$ itself should be analytically continued back to the Lorentzian regime at the end. Consequently, the Euclidean quantum-gravity path-integral is defined as

$$Z = \int_M D[g] e^{-I_E} , \qquad (21)$$

with the factor $\sqrt{-g}$ replaced by $-i\sqrt{g}$ and the Euclidean action $I_E = -iI$.

But the problem is not solved yet because the Euclidean gravitational action is not positive definite even for real positive definite metrics. Consider the situation of a compact manifold $M$ without boundary, so that there is no surface term in the action (which is to be the case for the remainder of the article) and perform a conformal transformation $g_{mn} \to \bar{g}_{mn} = \Omega^2 g_{mn}$ under which

$$R \to \bar{R} = \Omega^{-2} R - 6 \Omega^{-3} \Box \Omega . \qquad (22)$$

Then the action becomes

$$I_E[\bar{g}] = -\frac{1}{16\pi G}\int_M d^4x\sqrt{\bar{g}}(\Omega^2 R + 6\partial_m\Omega\partial_n\Omega \bar{g}^{mn} - 2\Lambda\Omega^4) \,. \tag{23}$$

One sees that $I_E[\bar{g}]$ can be arbitrarily negative by choosing a rapidly varying conformal factor $\Omega$. It has been shown how this problem can be circumvented in the case of the functional integral over all metrics which approach flat Euclidean space at infinity (asymptotically Euclidean metrics). The integral (21) can be decomposed into an integral over all such metrics with Ricci scalar $R = 0$, and an integral over the conformal deformations of these metrics. According to the *positive action theorem* the action of an asymptotically Euclidean metric with $R = 0$ is positive or zero, and vanishes if the metric is flat [15]. The troublesome behavior has therefore been isolated in the conformal degrees of freedom, and by rotating to an integration over complex conformal factors, which makes the kinetic term for the conformal factor positive, this can be rendered convergent also [16].

A second serious problem is connected to the fact that the coupling constant $G^{-1}$ has dimension of mass to the power $(n-2)$ and suggests that the theory is not perturbatively renormalizable above two dimensions.

## 2.2 Topology

There are attractive physical reasons for considering 4-geometries with different topologies. That is to say, there can occur quantum fluctuations of the metric not only within a given topology but from one topology to another suggesting a foam-like structure of spacetime on the scale of the Planck length [3]. This structure might give rise to an observable effect like the gravitational decay of baryons or muons. This can be caused, when they fall into gravitational instantons or virtual black holes and come out again as other species of particles. Evoked by this spacetime foam one could ask for quantum transition amplitudes between states specified by disconnected as well as connected 3-manifolds and multiply connected as well as simply connected ones. Unitarity would then suggest that the Euclidean functional integral contain a sum over topologically nontrivial 4-manifolds into which these 3-manifolds can be embedded.

One can imagine explicitly implementing this sum by first making a list of all physically distinct 4-manifolds, which are not diffeomorphic to each other. But it can be shown, that there is no classification scheme for $n$-topologies, $n \geq 4$, because there is no universal algorithm to decide when two entries on the list are the same manifold [17]. Subclasses are classifiable: if one imposes the condition that the manifolds be simply connected and admit a spin structure (vanishing $2^{nd}$ Stiefel-Whitney class) then they can presumably be classified (up to homotopy) by two topological invariants, the Euler characteristic $\chi$ and the Hirzebruch signature $\tau$. In the case of differentiable Riemannian manifolds, $\chi$ and $\tau$ can be expressed in terms of integrals involving the metric and the Riemannian

curvature tensor as

$$\chi = \frac{1}{128\pi^2} \int d^4x \sqrt{g} R_{abcd} R_{klmn} \epsilon^{abkl} \epsilon^{cdmn} \qquad (24)$$

$$\tau = \frac{1}{96\pi^2} \int d^4x \sqrt{g} R_{abcd} R^{ab}{}_{mn} \epsilon^{cdmn} . \qquad (25)$$

In a recent paper a wider class of manifolds, called conifolds were shown to be algorithmically decidable in four dimensions [18].

## 2.3 Measure

Several proposals for the nonperturbative integration measure have been suggested for quantum gravity. All of them are based on formal arguments which explain the variety of obtained results. Here, the origins of only a few are outlined, without any pretence of completeness.

The definition of the distance between two nearby geometries, whose difference $\delta g_{ij}$ is invariant under reparametrizations and ultralocal, is [6]

$$\|\delta g_{ij}\|^2 = \int d^4x d^4y \delta g_{ij}(x) G^{ijkl}(x,y) \delta g_{kl}(y) , \qquad (26)$$

where $G^{ijkl}(x,y)$ is the local bitensor density

$$G^{ijkl}(x,y) = \sqrt{g}(g^{ik}g^{jl} + g^{il}g^{jk} + \gamma g^{ij}g^{kl})\delta(x-y) \qquad (27)$$

and $\gamma \neq \frac{1}{2}$ but otherwise arbitrary. The determinant of the infinite continuous matrix $G^{ijkl}(x,y)$ can be computed at a formal level to get for the measure

$$D[g] = \prod_x G(x)^{\frac{1}{2}} \prod_{i \leq j} dg_{ij}(x) = \text{const} \prod_x g(x)^{\frac{(n-4)(n+1)}{8}} \prod_{i \leq j} dg_{ij}(x) , \qquad (28)$$

which in four dimensions, $n = 4$, is simply a constant.

Other results have been obtained demanding scale-invariance [19] and also in the Hamiltonian approach [20]

$$D[g] = \prod_x g(x)^{-\frac{n+1}{2}} \prod_{i \leq j} dg_{ij}(x) . \qquad (29)$$

To summarize, the discrepancies between various measures seem to lie in the definition of the product $\prod_x$ leading to different powers of $\sqrt{g}$ in the pre-factor. Although a rigorous definition of the measure on the space of geometries has been given allowing for an infinity of true Lebesgue measures the problem of which is the correct measure for quantum gravity is not yet solved at all [21].

# 3 Simplicial Quantum Gravity

Faced with these problems, various methods have been proposed to compute quantum-gravity path-integrals approximately. A well known way of doing so is to use simplicial manifolds leading to a lattice formulation of gravity called Regge calculus [10]. The Regge calculus needs a skeleton-like structure for the spacetime manifold and describes its geometry without using coordinates.

## 3.1 Simplicial manifolds

It is useful to begin by summarizing certain definitions and theorems on simplicial manifolds [18, 22].

Let $v_1, v_2 \ldots, v_{n+1}$ be affinely independent points[1] in $\mathbf{R}^{n+1}$. An $n$-simplex $s^n$ is the convex hull of these points:

$$s^n = \{x | x = \sum_{i=1}^{n+1} \lambda_i v_i; \ \lambda_i \geq 0; \ \sum_{i=1}^{n+1} \lambda_i = 1\} \ . \tag{30}$$

A 0-simplex is a point, a 1-simplex is a line segment or edge, a 2-simplex is a triangle, a 3-simplex is a tetrahedron. Higher dimensional simplices are generalizations of tetrahedra to higher dimensions. A simplex spanned by a subset of the vertices of an $n$-simplex is called a face. A simplex is uniquely determined by its vertices. This property is very important for both using and understanding simplicial complexes.

A simplicial complex $K$ is a topological space $|K|$ and a finite collection of simplices $s^p$ such that

1. $|K|$ is a closed subset of some finite dimensional Euclidean space,

2. if $s^p \in K$, then all faces of $s^p$ belong to $K$,

3. if $s^p, s^q \in K$, then either $s^p \cap s^q = \emptyset$ or $s^p \cap s^q$ is a common face of both $s^p$ and $s^q$.

The topological space $|K|$ is the union of all simplices in $K$. The dimension of a simplicial complex is the largest dimension of any simplex contained in the complex.

Thus, a simplicial complex describes both the building blocks of the space $|K|$ and gives the rules for how these building blocks are connected. Consequently the simplicial complex completely describes the topology of the space $|K|$. Finally, as each simplex in $K$ is uniquely determined by its vertices, the simplicial complex itself is uniquely determined by the vertices and the connection rules. It is clear that this property is especially valuable for computational purposes.

Simplicial complexes can describe topological spaces containing subspaces of different dimension, compact and noncompact spaces, and spaces with boundary.

---

[1] A set of points $\{v_i\}$ is affinely independent if the vectors $\{v_i - v_j\}$, $v_i \neq v_j$, are linearly independent for all $v_j$.

Thus, there are simplicial analogs of various standard definitions in topology. A pure simplicial complex is one in which every lower dimensional simplex is contained in at least one $n$-simplex where $n$ is the dimension of the simplicial complex. A compact simplicial complex is one which contains a finite number of simplices. A connected simplicial complex is one in which any two vertices are connected by a sequence of edges. A nonbranching simplicial complex is a simplicial complex of dimension $n$ in which every $(n-1)$-simplex is contained in at most two $n$-simplices.

At this point it is possible to define special sets of pure nonbranching simplicial complexes. They are given the name of pseudomanifolds. A pseudomanifold $P^n$ is a pure nonbranching simplicial complex such that any two $n$-simplices can be connected by a sequence of $n$-simplices, each intersecting along some $(n-1)$-simplex.

In order to study subsets of pseudomanifolds, two more definitions are required. These definitions are used to characterize the local topology of simplicial complexes and thus provide the means for defining simplicial equivalents of smooth manifolds. The star $St(v)$ of a vertex $v$ is the complex consisting of all simplices that contain $v$. The link $L(v)$ of a vertex is the subset of the star of $v$ consisting of all simplices in the star that do not intersect $v$ itself, cf. Figure 2.

Now the simplicial equivalents of smooth $n$-manifolds can be defined. A simplicial $n$-manifold is an $n$-pseudomanifold for which the link of every vertex is a simplicial $(n-1)$-sphere. A necessary and sufficient condition for the link to be a simplicial $(n-1)$-sphere is for the star of a vertex to be a simplicial $n$-ball. The main reason for defining simplicial manifolds in the above way is that it is a homogeneous definition, namely, no vertex of the simplicial complex has preferred treatment and the links of the vertices are all homeomorphic. This is similar to the idea of a topological manifold where each point has a neighborhood homeomorphic to a ball.

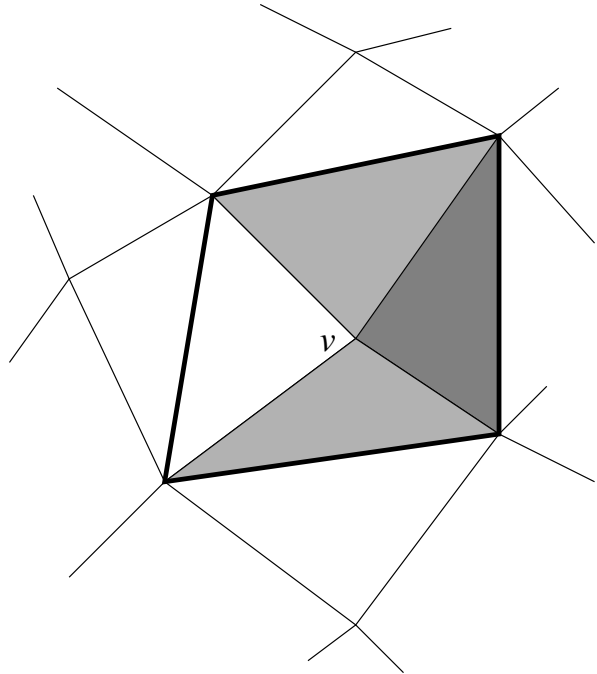

Figure 2: The star of a given vertex $v$ consists of the interior, edges, and vertices of those triangles which intersect $v$. The link of $v$ is composed of the heavily drawn edges and their vertices in the Figure. It is topologically equivalent to $S^1$.

In order to discuss the connection of smooth spaces to simplicial spaces, the first concept needed is that of a triangulation. A triangulation of a manifold $M^n$ consists of a simplicial manifold $K^n$ and a homeomorphism $t : |K^n| \to M^n$. Any topological space that admits a triangulation is homeomorphic to a polyhedron. Therefore, spaces allowing a triangulation are nice in the sense that they have the same properties as polyhedra. The most straightforward way to triangulate an $n$-dimensional manifold is to divide it into a sufficient number of hypercubes. Then draw the set of all diagonals (face, body, hyperbody diagonals) from each vertex to the other $2^n - 1$ points in the hypercube, i.e. the degrees of freedom per point are $2^n - 1$. Consequently the hypercube consists of $n!$ identical $n$-simplices and the coordination numbers are equal for each vertex. It is remarkable that in spite of the high symmetry the coordination numbers $N_m/N_n$ for the hypercubic triangulation are close to those of a random triangulation [23], which in general is desirable to use. Here $N_i$ denotes the number of $i$-simplices, $i = 0, \ldots, n$. A 2-dimensional example of this so-called regular triangulation is illustrated in Figure 3.

In four dimensions we have 24 4-simplices per hypercube [24], with $N_1 = 15N_0$, $N_2 = 50N_0$, $N_3 = 36N_0$, and $N_4 = 24N_0$. Thus, we have 15 edges $l_i$ per vertex, which can be labeled in the following way. Taking the initial point of some 15 edges as the origin of a Cartesian coordinate system, we assign the coordinates $(n_4, n_3, n_2, n_1)$, $n_i \in \{0, 1\}$, of the endpoint to each edge and indicate $l_i$ such that $i = n_1 + 2n_2 + 2^2 n_3 + 2^3 n_4$, cf. Table 1.

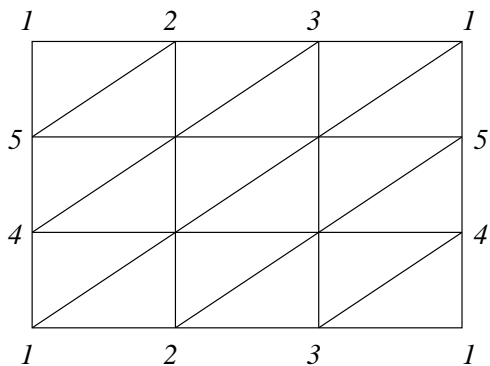

Figure 3: A regular triangulated 2-torus $T^2$ is represented as a rectangle with opposite sides identified.

| edges | face diagonals | body diagonals | hyperbody diagonals |
|---|---|---|---|
| 1 (0,0,0,1) | 3 (0,0,1,1) | 7 (0,1,1,1) | 15 (1,1,1,1) |
| 2 (0,0,1,0) | 5 (0,1,0,1) | 11 (1,0,1,1) | |
| 4 (0,1,0,0) | 6 (0,1,1,0) | 13 (1,1,0,1) | |
| 8 (1,0,0,0) | 9 (1,0,0,1) | 14 (1,1,1,0) | |
| | 10 (1,0,1,0) | | |
| | 12 (1,1,0,0) | | |

Table 1: Notation for labeling edges within a hypercube.

## 3.2 Regge action

To translate the path integral (21) into a concrete sum over simplicial histories it is necessary to be able to associate a metric and action with any $K^n$. Thus in order to proceed, metric information must be attached. The easiest way to do this is to require that the metric on the interior of each $n$-simplex is the Euclidean metric, that is all $n$-simplices in the simplicial complex are flat. With this requirement, the geometry of each $n$-simplex is completely fixed by specifying the lengths $l_i$ of all of its edges. Not every assignment of edge lengths is consistent with the simplices having flat interiors. The triangle inequalities and their analogs for tetrahedra and 4-simplices must be satisfied. In more mathematical terminology, the simplicial complex becomes a piecewise linear manifold, that in general does no longer fit in $\mathbf{R}^n$. It follows that the geometry of the simplicial complex is also completely fixed by specifying the lengths of all edges $l_i$ in the complex and one anticipates that all geometrical quantities such as volume and curvature can be expressed completely in terms of the edge lengths.

Indeed this is the case. Consider an $n$-simplex and define the $n$ vectors $e_i$ which start with the vertex 0 and proceed to the vertex $i$, i.e. the vectors $e_1, \ldots, e_n$ span the $n$-simplex. Its volume $V_n$ can be expressed by the generalization of Tartaglia's formula for a tetrahedron

$$V_n = \frac{1}{n!}\sqrt{\det(e_i \cdot e_j)}, \qquad (31)$$

where the $n \times n$-matrix of scalar products may be written in terms of the quadratic edge lengths $q_{ij} = |e_i - e_j|^2$ between vertices $i$ and $j$ by

$$e_i \cdot e_j = \frac{1}{2}(q_{0i} + q_{0j} - q_{ij}). \qquad (32)$$

Somewhat less obviously, curvature can also be expressed in terms of the edge lengths. As the metric on the interior of each $n$-simplex is flat, it is clear that the curvature of the combinatorial space is not carried on the interiors of the $n$-simplices. Rather, it turns out to be concentrated on the $(n-2)$-simplices of the simplicial complex. This becomes directly apparent in two dimensions, in which curvature is concentrated on vertices. The curvature is defined locally by the amount of rotation of a vector parallel-transported around an infinitesimally small loop. Since the space is flat in each triangle, the rotation of the vector is zero unless the loop contains a vertex $v$ inside. Then the circum-transported vector rotates by an angle $\delta(v)$. Such a region where the $\delta$-function-like curvature resides is called a hinge, and the rotation angle $\delta(v)$ is called deficit angle. Let $m$ denote the number of triangles in $St(v)$, then the deficit angle associated with this vertex is given by (cf. Figure 4)

$$\delta(v) = 2\pi - \sum_{i=1}^{m} \theta_i , \qquad (33)$$

and $\theta_i$ is the angle between two unit vectors that lie in adjacent edges of the $i^{th}$ triangle. Indeed the sum of the deficit angles over all vertices in the 2-manifold $M^2$ yields the Euler characteristic

$$\sum_{v \in M^2} \delta(v) = 2\pi \chi(M^2) , \qquad (34)$$

as required by the Gauss-Bonnet theorem.

The above discussion can be easily generalized for simplicial manifolds with dimension $n$. In this case the hinges are $(n-2)$-simplices. If the index $i = 1, \ldots, m$ sequentially labels adjacent $n$-simplices in $St(s^{n-2})$, then the deficit angle is given by

$$\delta(s^{n-2}) = 2\pi - \sum_{i=1}^{m} \theta_i , \qquad (35)$$

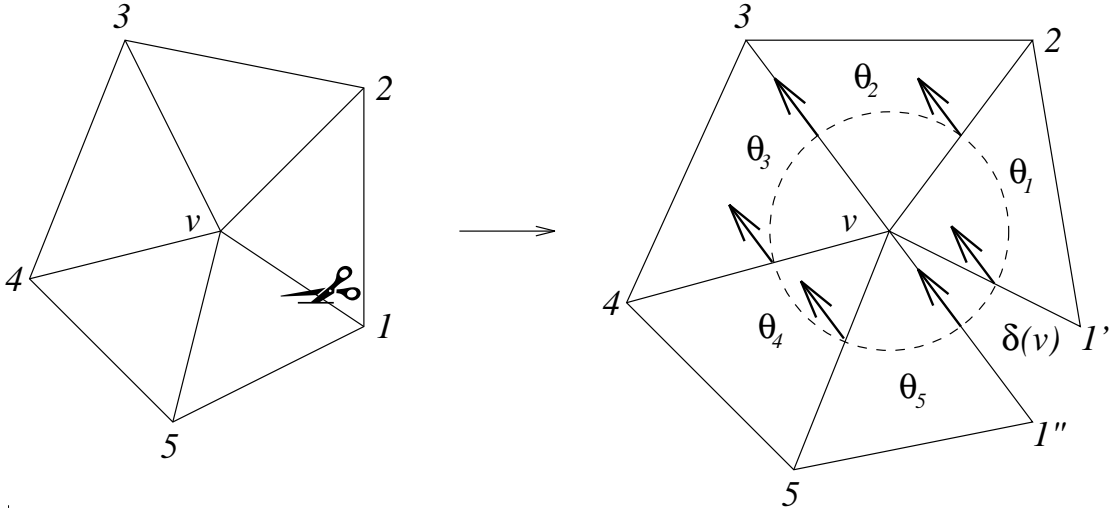

Figure 4: A 2-dimensional simplicial geometry is a net of flat triangles together with an assignment of lengths to their edges. The curvature is concentrated at the vertices and is measured by the deficit angle $\delta(v)$ which a parallel-transported vector experiences.

where $\theta_i$ is now the dihedral angle constructed from two unit vectors normal to $s^{n-2}$ that lie in the adjacent $(n-1)$-simplices of the $i^{th}$ $n$-simplex. The dihedral angle can be computed in terms of the edge lengths by elementary trigonometry in any dimension.

Finally, given the above definitions, the Regge action for Einstein gravity with cosmological constant $\Lambda$ for a closed pure nonbranching complex $K^n$ is [10, 11, 25]

$$I_R = -m_P^2 \sum_{s^{n-2} \in K^n} 2\delta(s^{n-2}) V(s^{n-2}) + m_P^2 \sum_{s^n \in K^n} 2\Lambda V(s^n) \;, \qquad (36)$$

where the first sum is over all $(n-2)$-simplices in the simplicial complex and the second is over all $n$-simplices in the complex. $V(s^n)$ is the $n$-volume of the indicated $n$-simplex and $m_P = (1/16\pi G)^{\frac{1}{2}}$ is the bare Planck mass. The Regge action for compact simplicial manifolds with boundary can also be formulated entirely in terms of edge lengths. Essentially one adds the appropriate discretized form of the boundary term (18) that appears in the continuum action [11]. A detailed analysis shows that in the classical limit the Regge action converges towards the Einstein-Hilbert action if the fatness (defined in the next Chapter) of each simplex is finite [26]. The classical continuum limit is reached with increasing the number of vertices $N_0 \to \infty$ and simultaneously decreasing the edge lengths $q_l \to 0$, i.e. the local lattice spacing becomes smaller than the local radius of curvature.

## 3.3 Group action for simplicial gravity

Given a simplicial lattice, to each edge of its dual lattice a Poincaré transformation can be assigned. With the new variables from these transformations, it is possible to build an action that reduces to the Regge action $I_R$ in the small curvature limit [27].

There is a general procedure for the construction of a dual lattice, whose cells are Voronoi polyhedra [23, 25]. The Voronoi polyhedron dual to vertex $P$ is the set of all points that are closer to $P$ than to any other vertex. In this way a $k$-simplex in the lattice is dual to an $(n-k)$-polyhedron in the dual lattice.

Let us consider a particular hinge $s^{n-2}$ with its vertices $P_1, \ldots, P_{n-1}$ and its star $St(s^{n-2})$ consisting of the set of $n$-simplices $s_1^n, \ldots, s_m^n$ characterized by $s_i^n = \{P_1, \ldots, P_{n-1}, Q_{i-1,i}, Q_{i,i+1}\}$, $(i = 1, \ldots, m)$. Further let us choose in each $n$-simplex $s_i^n$ a Lorentzian frame and denote the coordinates of the vertices of $s_i^n$ in this frame by

$$\begin{aligned} P_j &= \{y_j^a(i), \; a = 1, \ldots, n\} \\ Q_{i-1,i} &= \{z_{i-1,i}^a(i), \; a = 1, \ldots, n\} \;, \end{aligned} \qquad (37)$$

and also for each vertex of the dual lattice $D_i = \{x^a(i), \; a = 1, \ldots, n\}$ (cf. Figure 5). Furthermore, we associate to each edge of the dual lattice a Poincaré transformation

$$U(i, i+1) = \{U_b^a(i, i+1), U^a(i, i+1)\} \;, \qquad (38)$$

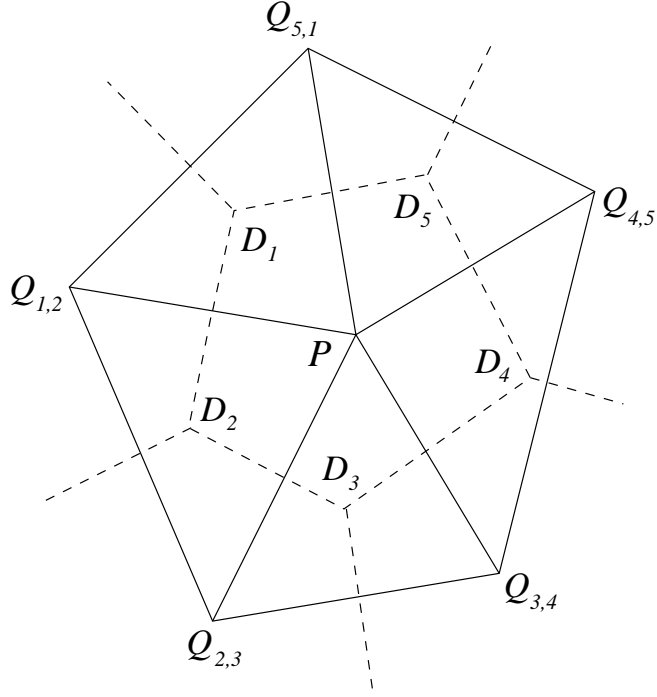

Figure 5: Part of a 2-dimensional lattice and its dual.

and define $U(i, i+1)$ by demanding

$$U^a_b(i, i+1) y^b_j(i+1) + U^a(i, i+1) = y^a_j(i) \tag{39}$$

and analogous for the other vertex coordinates. The arbitrariness of the choice of the reference frame in each $n$-simplex becomes in the dual lattice the gauge invariance under local Poincaré transformations $\Pi(i)$:

$$\begin{aligned} U(i, i+1) &\to \Pi(i) U(i, i+1) \Pi(i+1) \\ x^a(i) &\to \Pi^a_b(i) x^b(i) + \Pi^a(i) \ . \end{aligned} \tag{40}$$

By suitable fixing the translational part of the gauge it is always possible to set $x^a(i) = 0 \ \forall i$. Therefore, the plaquette variable on the dual lattice around the hinge $s^{n-2}$

$$W_i(s^{n-2}) = U(i, i+1) U(i+1, i+2) \ldots U(i-1, i) \tag{41}$$

leaves the coordinates of all vertices $P_j$ of $s^{n-2}$ unchanged and becomes a rotation by the deficit angle $\delta(s^{n-2})$ in a plane perpendicular to $s^{n-2}$.

This leads to the following proposal for a "compact" gravitational action on the lattice in terms of plaquette variables

$$\begin{aligned} I_C &= -m_P^2 \sum_{s^{n-2}} W_i^{a_1 a_2}(s^{n-2}) \prod_{j=1}^{n-2} [y_j^{a_{j+2}}(i) - y_{n-1}^{a_{j+2}}(i)] \epsilon_{a_1 \ldots a_n} \\ &= -m_P^2 \sum_{s^{n-2}} \sin \delta(s^{n-2}) V(s^{n-2}) \ . \end{aligned} \tag{42}$$

It is obvious that (42) reduces to (36) in the limit of small deficit angles, further it bears a striking resemblance with the continuum action written in terms of differential forms.

In deriving (42) the Levi-Civita connection has been implicitely used for the parallel transport within a loop. But $I_C$ can also be defined for general connections, including ones with nonvanishing torsion and even ones not determined by the metric [28].

One disadvantage of the compact action is, that it does not have the property $I_R(M^n) = I_R(M_1^n) + I_R(M_2^n)$ like the Regge action, whereby we have defined $M^n = M_1^n + M_2^n$ as the manifold obtained by joining together the manifolds $M_1^n$ and $M_2^n$ along $M^{n-1}$. It is therefore not clear, whether it gives rise to a unitary theory in the form of physical- or reflection positivity [28].

## 3.4 Simplicial path integral

Given a method of associating a geometry and an action with any complex $K^n$ the Regge equivalent of the path integral (21) is

$$Z = \sum_{K^n} \int D[q] e^{-I[q]} , \qquad (43)$$

where the integration over all $n$-geometries without boundary has been replaced by a sum over distinct simplicial complexes $K^n$ and integrating over their quadratic edge lengths $q$. Usually one abandons summing over different $K^n$ but contents oneself with only one fixed toplogy. Anyhow, arguments have been pointed out that a sum over topologies cannot give a finite functional integral, because the number of complexes within a given topology increases too rapidly as a function of the number of simplices [28].

Two somehow complementary methods have been established to evaluate the simplicial path integral.

a) **The Regge approach:** The triangulation is kept fixed and the edge lengths, which play the role of infinitesimal geodesics in the continuum are allowed to vary. The path integral (43) reduces to a summation over different configurations $[q]$ [11, 25, 29, 30].

b) **The dynamical triangulation approach:** This method represents an alternative and complementary approach to what will be discussed here [31]. The simplicial manifold consisting of a set of equilateral simplices is locally changed by applying elementary moves (e.g. the well-known Alexander moves [32]) on the incidence matrices, i.e. one performs a summation over all possible triangulations of the manifold. In two dimensions this method is equivalent to matrix models, but in four dimensions problems seem to appear, because it is not possible to triangulate flat space with equilateral 4-simplices and therefore to reach the classical continuum limit directly.

Since within the Regge approach the squares of the edge lengths, $q$, are linearly related to $g_{mn}$, the corresponding measure to $D[g]$ would be $D[q] = \prod_{q_l} d\mu(q_l)$ indicating the freedom available in choosing the measure on the space of edge lengths. Investigations concerning regularly triangulated hypercubic lattices suggest [24, 25, 26, 29, 30]

$$D[q] = \prod_l q_l^{\sigma-1} dq_l \mathcal{F}\{q\} \ , \tag{44}$$

with $\mathcal{F} = 1$ if the generalized triangle inequalities are fulfilled and $\mathcal{F} = 0$ otherwise. The parameter $\sigma \geq 0$ determines the behavior of the measure under rescaling. The value $\sigma = 1$ corresponds to the uniform measure (28), whereas for $\sigma = 0$ a scale-invariant measure results (29).

The functional integral over metrics has now been reduced to the product of integrals over edge lengths as the metric is discretized. This has the consequence that diffeomorphism invariance is destroyed. The recovery of the spacetime diffeomorphism group is of vital significance in discussing the continuum limit and can be approached from several perspectives. For example, a triangulation of a manifold $M$ can be viewed as as a continuous map $\tau$ from an abstract simplicial complex $K$ to $M$. Then each diffeomorphism $\phi$ of $M$ generates a new triangulation $\phi \circ \tau : K \to M$. Unfortunately, the discretized Regge action is not invariant under the change in edge lengths, there remains only an approximate invariance group, which becomes increasingly accurate with increasing refinement of the lattice. This is revealed by the fact that the Hessian matrix constructed out of second derivatives of the action with respect to edge lengths is generally nonsingular [11] (unlike the continuum counterpart). Whether or not such a structure would suffice for quantization is not known. But in any event, there is a clear danger of overcounting the metric modes and thus a type of gauge fixing for the triangulation was developed [33]. Given a complex $K$, to each metric on $M$ corresponds just one triangulation $\tau$ obtained by minimizing the deviations of the edge lengths from their mean. This is realized by dynamical triangulation, which is plagued by other problems like ergodicity of the path integral [34].

# 4 Entropy-dominated Phase

Different formulations of the simplicial action lead to the same classical continuum limit, but it is not clear to what extent they differ after quantization. Unfortunately, the computation of the path integral is not straightforward, because the action is still unbounded and the Regge calculus admits no simple means of dealing with conformal modes and hence controlling this deficiency. Indeed, both the original Regge action (36) in four dimensions, $n = 4$,

$$I_R = -\beta \sum_t A_t \delta_t + \lambda \sum_s V_s \tag{45}$$

as the action (42) with the compact deficit angle contribution and an additional cosmological term

$$I_C = -\beta \sum_t A_t \sin \delta_t + \lambda \sum_s V_s \tag{46}$$

can be made arbitrarily negative if some of the 4-simplices have near-zero 4-volume $V_s$ but contain very-large-area triangles $A_t$, whose deficit angles are positive. It is reasonable to set an appropriate cut-off via a lower limit $f$ on the fatness

$$\phi_s = C^2 \frac{V_s^2}{\max_{l \in s}(q_l^4)} \,, \quad C = 24 \,, \tag{47}$$

of each 4-simplex, $\phi_s \geq f > 0$ [26]. The fatness is maximal for equilateral simplices and goes to zero for collapsing ones.

But even without any cut-off the unboundedness of the action need not to render the path integral ill-defined. The reason is that entropy effects coming from the measure could cure this pathology [29]. This can be seen if the path integral is rewritten as

$$Z = \int_{-\infty}^{\infty} dI \, n(I) e^{-I} \,, \tag{48}$$

where $n(I)$ gives the density of states for the action $I$. If $n(I)$ goes faster to zero with $I \to -\infty$ than the exponential diverges, then $Z$ would exist. Numerical computations indicate that the entropy of the system can indeed compensate the unbounded action [25, 29, 30].

## 4.1 Monte Carlo method

A possible nonperturbative method to approximate expectation values of observables $\mathcal{O}$

$$\langle \mathcal{O} \rangle = \frac{1}{Z} \int \prod_l d\mu(q_l) \mathcal{O}(\{q_l\}) e^{-I(\{q_l\})} \tag{49}$$

is by using numerical Monte Carlo (MC) techniques that do not rely on an expansion in a small parameter. A finite discrete system is considered, which is no serious restriction from the viewpoint of computer simulations, because a continuous variable is anyhow discretized by the computer's digital accuracy. For a finite system, (49) is replaced by

$$\langle \mathcal{O} \rangle = \frac{1}{Z} \sum_k \mathcal{O}_k e^{-I_k} , \qquad (50)$$

where the possible configurations $\{q_l\}_k$ are labeled by integers $k$ and therefore $I_k = I(\{q_l\}_k)$. The division by the partition function $Z$ is to normalize expectation values, such that $\langle \mathbf{1} \rangle = 1$. Since the total number of configurations is exorbitantly large one performs an importance sampling, so that configurations are generated with probability

$$P(\{q_l\}_k) = \text{const } e^{-I_k} . \qquad (51)$$

The constant is determined by the normalization condition $\sum_k P(\{q_l\}_k) = 1$. Configurations with (51) as their equilibrium distribution can be produced by a Markov process. The elements of the Markov chain are the configurations. They are sequentially generated such that $\{q_l\}_j$ depends only on the previous configuration $\{q_l\}_{j-1}$. The transition probability to create $\{q_l\}_j$ from $\{q_l\}_i$ is given by the matrix element $W_{ij} = W(i \to j) \geq 0$, which is required to satisfy the following properties:

**i) Ergodicity:**

$$e^{-I_i} > 0 , \quad e^{-I_j} > 0 \qquad (52)$$

implies the existence of an integer number $m > 0$, such that $(W^m)_{ij} > 0$.

**ii) Normalization:**

$$\sum_j W_{ij} = 1 . \qquad (53)$$

**iii) Balance:**

$$\sum_i W_{ij} e^{-I_i} = e^{-I_j} \qquad (54)$$

means, the Boltzmann weight factors form eigenvectors of the matrix $W$.

An ensemble is a collection of configurations such that to each configuration a probability $P_i$ is assigned and $\sum_i P_i = 1$. The Boltzmann ensemble $E^B$ is defined to be the ensemble with probability distribution

$$P_i^B = \text{const } e^{-I_i} , \quad \sum_i P_i^B = 1 . \qquad (55)$$

An equilibrium ensemble $E_e$ of the Markov process is defined by satisfying $WP_e = P_e$. The central point for the MC method is, that under the conditions i), ii), and iii) the Boltzmann ensemble is the only equilibrium ensemble of the Markov process (see [29] for a proof). There are many ways to construct a Markov process satisfying these conditions. In practise most MC algorithms are based on the condition of *detailed balance*:

$$W_{ij} e^{-I_i} = W_{ji} e^{-I_j} . \qquad (56)$$

Using the normalization condition ii) $\sum_j W_{ij} = 1$, detailed balance immediately implies balance iii). A popular choice is the Metropolis algorithm [35] because of its computational simplicity. This procedure is described below for the Regge skeleton.

A new configuration with $I_k$ is proposed by varying the edge lengths of the lattice by a small amount according to the measure (44). If it is not in the Euclidean sector, i.e. the triangle inequalities are not fulfilled, or if the fatness $\phi_s$ is lower than the limit $f$, it is rejected. Otherwise one has to check the Metropolis condition: the new configuration $I_k$ is always accepted for $I_k < I_j$ and for $I_k > I_j$ it is accepted with probability $\exp(I_j - I_k)$.

A MC iteration is finished, when a new value is proposed for every edge in the lattice. In the limit of an infinite number of iterations the expectation values are given by

$$\langle \mathcal{O} \rangle = \lim_{T \to \infty} \frac{1}{T} \sum_{k=1}^{T} \mathcal{O}_k(\{q_l\}) \ . \tag{57}$$

In practice, the generated configurations are averaged over a finite number of iterations after thermalization to approximate the considered expectation value.

## 4.2 Results

MC simulations have been performed to investigate the entropy-dominated phase for both proposed simplicial gravitational actions, (45) and (46), with the following features and parameters.

**Lattice:** The regularly hypercubic triangulated simplicial manifold is fixed to be homeomorphic to a 4-torus $T^4$. The total number of vertices is $N_0 = 3^3 \times 8$.

**Fatness:** To facilitate the simulations a lower limit $f = 10^{-4}$ is set for the fatness $\phi_s$ of each 4-simplex. This scale invariant cut-off reduces the configuration space and thus allows to reach an equilibrium after a reasonable number of MC iterations. Previous computations suggest that such a restriction does not affect the behavior of the system in the well-defined phase [30].

**Measure:** Investigations of a 1-parameter family of measures (44) indicate the stability of the entropy-dominated region against variations of the measure [30]. As the simplest choice the uniform measure with $\sigma = 1$ is used, but universality of the results is expected.

**Couplings:** The bare cosmological constant appearing in the gravitational action (45) or (46) is set to $\lambda = 1$ by fixing the average lattice volume. $\beta$ determines the gravitational coupling and is varied in the range $[0.07, 0.15]$.

**Iterations:** 100k - 300k MC sweeps inclusive thermalization have been applied for each value of $\beta$.

Examining the scaling properties of the action and the measure, a very useful identity for checking the accuracy of numerical computations is obtained [25]. The path integral

$$Z(\beta, \lambda) = \int \prod_l dq_l \, q_l^{\sigma-1} \mathcal{F}\{q_l\} \exp(\beta \sum_t A_t \delta_t - \lambda \sum_s V_s) \tag{58}$$

obeys

$$Z(\beta, \lambda) = \left(\frac{\beta}{\lambda}\right)^{\sigma N_1} Z\left(\frac{\beta^2}{\lambda}, \frac{\beta^2}{\lambda}\right), \tag{59}$$

which implies

$$2\lambda \langle \sum_s V_s \rangle - \beta \langle \sum_t A_t \delta_t \rangle = \sigma N_1, \tag{60}$$

an equation easily to verify in simulations. The correspondence between analytical and numerical results is shown in Figure 6 for the Regge action. The same behavior is found for the compact action $I_C$.

A quantity of physical interest is the average curvature

$$\mathcal{R} = \frac{\langle \sum_t 2 A_t \delta_t \rangle}{\langle \sum_s V_s \rangle} \langle q \rangle \sim \frac{\int \sqrt{g} R}{\int \sqrt{g}}, \tag{61}$$

which can be understood as effective cosmological constant [30]. The variables $A_t, V_s, q$ are expressed in units of the bare Planck length.

The behavior of the average curvature as a function of the coupling $\beta$ is depicted in Figure 7. As $\beta$ is varied a phase transition is found separating a smooth from a rough phase of gravity. For sufficiently small $\beta$ the curvature is small and negative and enters a region of large curvature for $\beta > \beta_{crit}$. How far the location of the transition point depends on the choice of the fatness, the measure, and the underlying triangulation is discussed in [30]. While the transition is rather discontinuous for the Regge action $I_R$, it is smooth for $I_C$, which might be important for the continuum limit. This suggests that the appearance of $\sin \delta_t$ in the action provides a dynamical mechanism to force the curvature to be small.

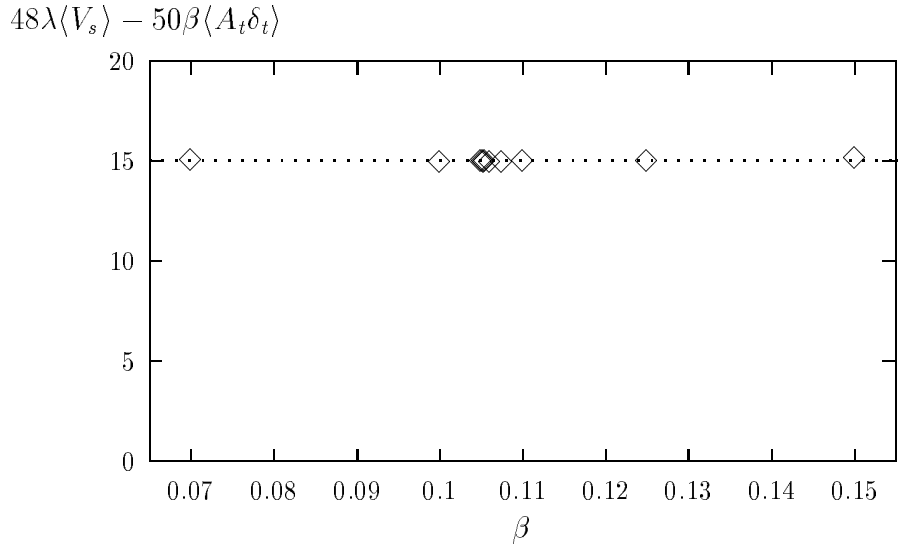

Figure 6: For the $3^3 \times 8$ lattice with $\lambda = \sigma = 1$ one has $\frac{\sigma N_1}{N_0} = 15$, which is in perfect agreement with our simulation data, here exemplified for the Regge action $I_R$. Error bars are smaller than the symbol size.

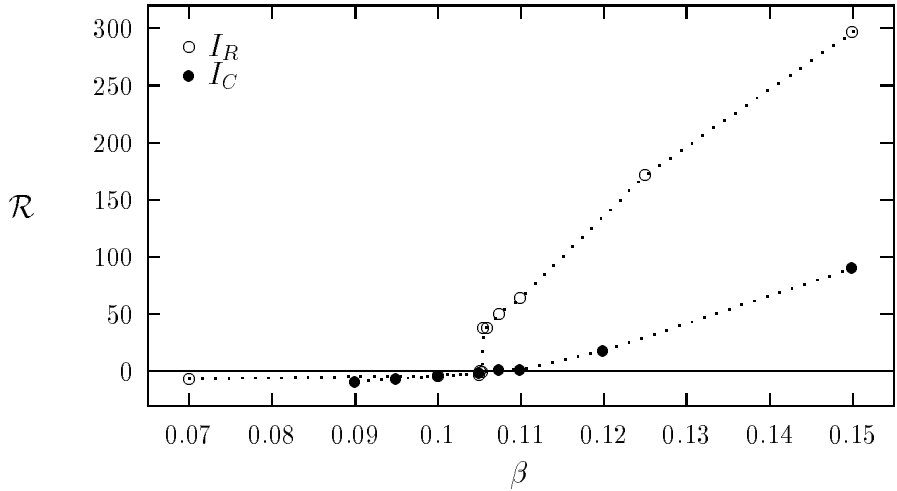

Figure 7: Average curvature using the Regge and the compact action $I_R$ and $I_C$, respectively, on a $3^3 \times 8$ lattice with $\lambda = \sigma = 1$ for different gravitational couplings $\beta$. In the well-defined phase most data points from both actions overlap. Error bars are smaller than the symbol size.

A similar effect was reported when higher-order terms in the curvature, e.g. $R^2$ terms, are added to the Regge action $I_R$ and simulated within the uniform measure. It was stated that the phase transition to the ill-defined region suggests some sort of multicritical behavior with a line of $1^{st}$-order transitions without any higher-order term and a line of $2^{nd}$-order transitions with $R^2$ terms [25]. Unfortunately, this advantage for a continuum limit is accompanied by a violation of unitarity. Originally, the Einstein Lagrangian was replaced with one that involved squares of the Riemannian curvature because of perturbative renormalizability. But this is achieved at the expense of acquiring a massive spin-2 ghost partner (*Poltergeist*) for the physical massless graviton [36].

The transition to the region of large curvature is accompanied by the tendency of the simplices to collapse into degenerate configurations. This is seen if one examines the expectation value of the fatness $\langle \phi_s \rangle$. Figure 8 shows $\langle \phi_s \rangle$ as a function of the coupling $\beta$ for both actions. It turns out that in the well-defined phase the average fatness is much larger than the lower limit $f = 10^{-4}$. With the transition to the ill-defined phase the fatness decreases (suddenly) signaling the presence of collapsed 4-simplices (spikes) and a crumpled lattice with fractal dimension less than four. This region of the phase diagram can only be observed for finite $f$, because without any cut-off the action diverges and simulations would not reach an equilibrium.

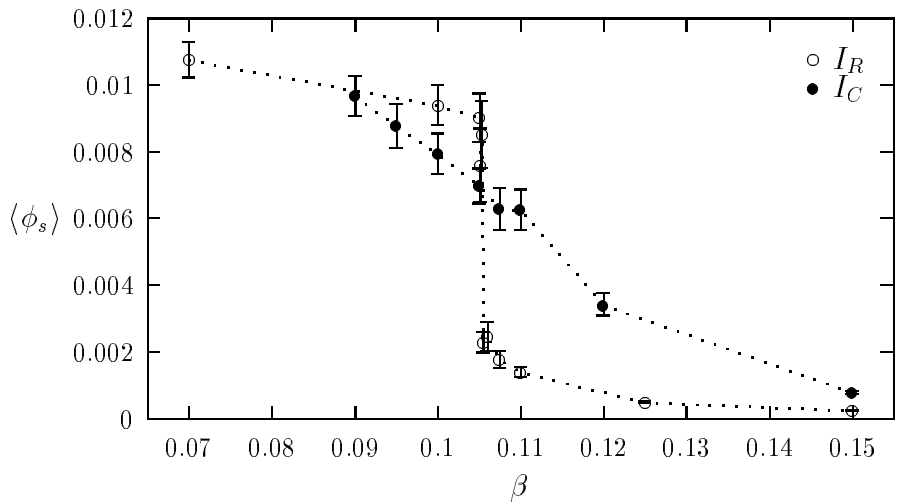

Figure 8: Fatness $\langle \phi_s \rangle$ versus $\beta$ in the $3^3 \times 8$ lattice with $\lambda = \sigma = 1$ for the Regge action and the compact action. The cut-off for the fatness was set to $f = 10^{-4}$ to allow equilibrium in the ill-defined phase.

The development of spikes is reflected in the formation of an inhomogeneous distribution of the edge lengths. Figure 9 depicts the histograms for a configuration of edge lengths below and above the transition point using the compact action $I_C$ (46). While a homogeneous distribution is typical for the well-defined phase, it is rather inhomogeneous in the ill-defined regime. A similar behavior for the Regge action $I_R$ (45) was found in [30].

To summarize, it seems that the system prefers configurations with distributions of edge lengths in the well-defined phase leading to a small negative average curvature. Increasing the coupling $\beta$ one reaches a critical value entering into the ill-defined phase characterized by degenerate configurations with collapsing simplices and very large positive curvatures because of the unbounded action. Note that this collapsed phase is the coupling region of the weak-field expansion $G \to 0$.

In principle it could be possible that the well-defined region of the path integral is only a numerical artefact due to metastable states. Numerical simulations examining inhomogeneous start configurations with very large positive curvature, however, indicate independence from the initial conditions, i.e. the average curvature returns to small and negative values for coupling parameters of the well-defined phase [30].

Another important task is to study the influence of the underlying triangulation on the entropy-dominated phase. The most natural setting would be a random lattice reproducing spacetime isotropy stochastically. Since random lat-tice simulations are technically cumbersome, most numerical computations have been performed on the regular hypercubic triangulation of the 4-torus. Nevertheless, to investigate the effect of the lattice structure one can apply barycentric subdivisions and study such irregularly triangulated systems as the number of randomly inserted vertices increases. Recent computations reveal that the form of the incidence matrix and the local coordination number have an important influence on the critical coupling [37].

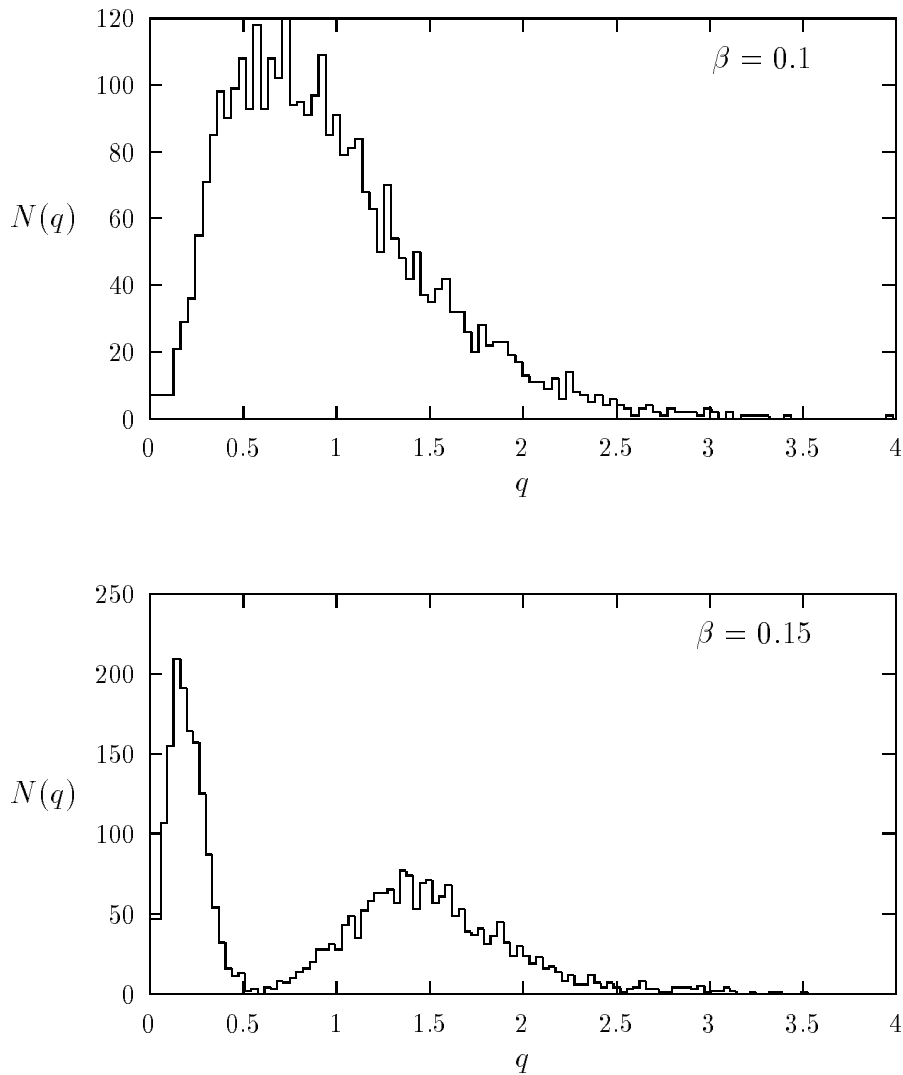

Figure 9: Histograms of the squared edge lengths $q$ in units of the average squared edge length $\bar{q} = \frac{1}{N_1} \sum_l q_l$ of a typical configuration below (upper picture) and above (lower picture) the phase transition applying the compact action $I_C$.

# 5 Two-point Functions in Simplicial Quantum Gravity

The previous discussion has dealt exclusively with averages of local operators, which provide a considerable amount of information. To understand the interaction mechanism of a theory one has to study correlation functions, which yield the mass spectrum of the system. This can be used to decide whether a massless graviton does exist or not.

## 5.1 Weak-field limit

For a 4-dimensional manifold, regularly triangulated as described in Section 3.1, the variation of the action (45) ($\lambda = 0$) with respect to the edge length $l$ is

$$\delta I_R = \delta(\sum_t A_t \delta_t) = \sum_t \delta A_t \, \delta_t = \sum_t \frac{\partial A_t}{\partial l} \delta l \, \delta_t \; , \qquad (62)$$

using the fact that $\sum_t A_t \, \delta \delta_t = 0$ [10]. Hence the equations of motion are

$$\sum_t \frac{\partial A_t}{\partial l} \delta_t = 0 \; , \qquad (63)$$

being the lattice analog of the Einstein equations. To derive the free two-point function the second variation of $I_R$ around flat space, where the deficit angles all vanish, yields

$$\begin{aligned} \delta^2 I_R &= \sum_t (\delta^2 A_t \delta_t + \delta A_t \, \delta \delta_t) \\ &\cong \sum_t (\delta A_t \, \delta \delta_t) \\ &= \sum_t (\sum_l \frac{\partial A_t}{\partial l} \delta l)(\sum_l \frac{\partial \delta_t}{\partial l} \delta l) \; . \end{aligned} \qquad (64)$$

Now let $\delta_i$ ($i = 1, \ldots, 15$) be the amount of fluctuation of $l_i$

$$\delta l_i = \delta_i l_i \; , \qquad (65)$$

then one can write (64) in the bilinear form

$$\delta^2 I_R = \delta^\dagger M \delta \; , \qquad (66)$$

where $\delta$ is a column vector with 15 components per lattice point and the matrix $M$ corresponds to the inverse propagator.

Expanding the fluctuation $\delta_i^{(a,b,c,d)}$ at a point with coordinates $(a, b, c, d)$ into periodic modes $\omega_i = \exp(\frac{2\pi i}{n_i})$ ($i = 1, 2, 4, 8$)

$$\delta_i^{(a,b,c,d)} = (\omega_1)^a (\omega_2)^b (\omega_4)^c (\omega_8)^d \delta_i \; , \qquad (67)$$

one can write (66) in momentum space

$$(\delta^2 I_R)_\omega = \delta_i^\dagger M_\omega^{ij} \delta_j \ . \tag{68}$$

By a unimodular similarity transformation $\delta \to \delta'$ one can block diagonalize $M_\omega$. It turns out that five modes are completely decoupled from the other ten and do not affect the dynamics of the system. Therefore, one considers only those parts of the vector $\delta'$, whose degrees of freedom have the same multiplicity of ten as $g_{mn}$ in the continuum theory.

By means of a further transformation $\delta' \to \bar{h}$ the dynamical part of (68) is

$$\frac{1}{2} L_R = \bar{h}^\dagger 2 \sum_{i=1,2,4,8} (1 - \cos \frac{2\pi}{n_i}) \begin{pmatrix} \frac{1}{2}B & 0 \\ 0 & I_6 \end{pmatrix} \bar{h} - \bar{h}^\dagger C^\dagger C \bar{h} \ . \tag{69}$$

$B$ is a $4 \times 4$ matrix, $I_6$ the 6-dimensional identity matrix and $C = C(\omega)$ a $4 \times 10$ matrix. For a concrete expression see [24].

In the continuum theory the Langrangian of the weak gravitational field $h_{kl}$ with gauge breaking term $C_m$ can be written as [40]

$$\mathcal{L} = -\frac{1}{2} \partial_a h_{kl} V_{klmn} \partial_a h_{mn} + \frac{1}{2} C_m^2 \ , \tag{70}$$

where

$$V_{klmn} = \frac{1}{2} \delta_{km} \delta_{ln} - \frac{1}{4} \delta_{kl} \delta_{mn} \tag{71}$$

$$C_m = \partial_n h_{mn} - \frac{1}{2} \partial_m h_{nn} \ . \tag{72}$$

To compare this expression with the result in the simplicial case, trace reversed variables are first employed

$$\bar{h}_{mn} = h_{mn} - \frac{1}{2} \delta_{mn} h_{ll} \ , \tag{73}$$

and then the fluctuation variables relabeled such as

$$\begin{aligned} \bar{h}_m &= \bar{h}_{mm} \\ \bar{h}_{m+n} &= \bar{h}_{mn} \ , \ m \neq n \ . \end{aligned} \tag{74}$$

Translating the variables into momentum space one finds

$$L = \frac{1}{2} k_m \bar{h}^\dagger \begin{pmatrix} \frac{1}{2}B & 0 \\ 0 & I_6 \end{pmatrix} k_m \bar{h} - \frac{1}{2} \bar{h}^\dagger \hat{C}^\dagger \hat{C} \bar{h} \ , \tag{75}$$

where $k_m = i \partial_m$ is the momentum and $\hat{C} = \hat{C}(k)$. By setting $k_i = -\frac{2\pi}{n_i}$, in the long wave length limit one can expand

$$\omega_i = 1 - i k_i - \frac{1}{2} k_i^2 + \dots \ , \tag{76}$$

and gets

$$2 \sum_{i=1,2,4,8} (1 - \cos \frac{2\pi}{n_i}) \to \sum_{i=1,2,4,8} k_i^2 \ , \ C \to \hat{C} \ . \tag{77}$$

Inserting this in Equation (75) and comparing with (69) the continuum and lattice correspondence is clear (up to a trivial normalization factor) in the case of a weak gravitational field.

## 5.2 Correlation functions

The continuum weak-field propagator can be formulated in terms of spin-2 and spin-0 projection operators [41]

$$\langle h_{kl}(x) h_{mn}(y) \rangle \sim \int d^4p \frac{P^{(2)}_{klmn} - 2P^{(0)}_{klmn}}{p^2} e^{-ip(x-y)} \; , \tag{78}$$

with

$$P^{(2)}_{klmn} = \frac{1}{3}\left(\frac{p_k p_l}{p^2}\delta_{mn} + \frac{p_m p_n}{p^2}\delta_{kl}\right) -$$
$$- \frac{1}{2p^2}(p_k p_m \delta_{ln} + p_k p_n \delta_{lm} + p_l p_m \delta_{kn} + p_l p_n \delta_{km}) +$$
$$+ \frac{1}{2}(\delta_{km}\delta_{ln} + \delta_{kn}\delta_{lm}) - \frac{1}{3}\delta_{kl}\delta_{mn} + \frac{2}{3p^4} p_k p_l p_m p_n \; , \tag{79}$$

$$P^{(0)}_{klmn} = -\frac{1}{3}\left(\frac{p_k p_l}{p^2}\delta_{mn} + \frac{p_m p_n}{p^2}\delta_{kl}\right) + \frac{1}{3}\delta_{kl}\delta_{mn} + \frac{1}{3p^4} p_k p_l p_m p_n \; . \tag{80}$$

To probe the correlations in the spin-0 and spin-2 part separately one uses the fact that fluctuations in the volume are sensitive to the scalar channel

$$\langle \sqrt{g}(x) \sqrt{g}(y) \rangle \sim - \int d^4p \frac{P^{(0)}}{p^2} e^{-ip(x-y)} \; , \tag{81}$$

whereas fluctuations in the curvature are due to presence of spin-2 particles

$$\langle \sqrt{g} R(x) \sqrt{g} R(y) \rangle \sim \int d^4p\, e^{-ip(x-y)} \int d^4q \tilde{R}(p,q) \frac{P^{(2)}}{p^2} \tilde{R}(p,q) \frac{P^{(2)}}{(p-q)^2} \; , \tag{82}$$

where $\tilde{R}$ represents some momentum dependent vertex [42].

If $\lambda \neq 0$, the cosmological term implies a contribution to the propagator and the appearance of a tadpole diagram. With the gauge breaking term (72) the Langrangian becomes

$$\mathcal{L} = \lambda(1 + \frac{1}{2}h_{ll}) + \frac{1}{2}h_{kl} V_{klmn}(\partial^2 - \lambda) h_{mn} \; , \tag{83}$$

leading to almost the same propagator as before (78), except that $p^2$ is replaced by $p^2 + \lambda$ in the denominator. This propagator therefore corresponds to the exchange of a particle with mass $m = \sqrt{\lambda}$. At first instance the gravitational force becomes a Yukawa force with a range given by this mass. But it has been conjectured that by expanding around the correct solution of the classical equations of motion this mass term disappears, because in the presence of a cosmological term flat space is no longer a solution of Einstein's equations. Up to now this has not been worked out even for simple cases [40].

In case of the Regge approach connected correlation functions

$$G(d) = \langle \mathcal{O}(x) \mathcal{O}(y) \rangle_c \tag{84}$$

of a local operator $\mathcal{O}$ are measured at two points separated by the geodesic distance $d = |x - y|$ [42]. The geodesic distance $d$ between two points $x$ and $y$ in a discrete space is the length of the shortest path joining them.

Since the edge lengths are the dynamical variables of the theory a logical analog of (78) would be the $15 \times 15$ matrix

$$G^{(q)}_{ab}(d) = \langle q_a(d) q_b(0) \rangle_c . \tag{85}$$

The indices $a, b$ label the different types of edges within a hypercube. The squared edge lengths are used mainly for technical reasons. Some care must be taken to interpret (85) as a propagator. From perturbative calculations considering small fluctuations around flat space (cf. previous Section) it is already known that the free inverse propagator on the lattice has zero eigenvalues. One can argue that these zero modes are not present on a general curved lattice since diffeomorphism invariance is lost. Moreover, the formulation of gauge transformations and gauge fixing is merely worked out for lattice gravity because the Regge calculus admits no natural means of introducing a ghost determinant or a gauge breaking term. But the fact that the lattice theory is already formulated in a coordinate invariant manner at least legitimates correlations of the type (85). Also correlations between deficit angles $G^{(\delta)}_{ab}(d) = \langle \delta_a(d) \delta_b(0) \rangle_c$, with the indices $a, b$ labeling the 50 triangles attached to a vertex within one hypercube, are confronted with the above problems, but can give insight into the interaction mechanism of the system.

The simplicial analogs of the invariant two-point functions (81) and (82) correspond to the volume correlations

$$G^{(V)}(d) = \langle \sum_{s \supset v_0} V_s \sum_{s' \supset v_d} V_{s'} \rangle_c \tag{86}$$

and the curvature correlations

$$G^{(R)}(d) = \langle \sum_{t \supset v_0} \delta_t A_t \sum_{t' \supset v_d} \delta_{t'} A_{t'} \rangle_c . \tag{87}$$

Here $\sum_{s \supset v_0} V_s$ denotes the sum over all volumes of 4-simplices meeting at the vertex $v_0$ and $\sum_{t \supset v_0} \delta_t A_t$ collects the contributions of all triangles at $v_0$. The vertices $v_0$ and $v_d$ are separated by the geodesic distance $d$.

## 5.3 Results

MC simulations have been performed to compute several correlation functions for both simplicial gravitational actions, (45) and (46). In the presentations the index labeling the vertices in the long lattice direction is taken as a number for the actual distance, henceforth called *index distance*, between local operators. This seems to be reasonable because in the well-defined phase with its small average curvature the index distance is presumably a good approximation of the true geodesic distance. Further, it avoids a time consuming determination of geodesic distances via random walk on the fluctuating simplicial lattice.

**Lattice:** A hypercubic triangulated 4-torus with $N_0 = 3^3 \times 8$ and $N_0 = 4^3 \times 16$ vertices is used. Because of the inherent periodicity of the 4-torus the maximum index distance for two-point functions is 4 and 8, respectively.

**Fatness:** The fatness cut-off has been set to $f = 10^{-4}$ and allows the study of correlations even for gravitational couplings $\beta > \beta_c$.

**Gravitational coupling:** $\beta$ is varied within the range $[0.07, 0.15]$ for the small lattice and $[0.1, 0.14]$ for the large.

**Measure and cosmological constant:** They are set equal, $\sigma = \lambda$, so that $\langle V_s \rangle$ takes the same value $N_1/2N_4$ for $\sigma > 0$ at $\beta = 0$, cf. (60). In particular, two values of $\sigma = \lambda$ were chosen for every lattice size: 0.25 and 1.0 for the smaller and 0.1 and 1.0 for the larger.

**Iterations:** 100k - 300k and 50k MC sweeps were produced for the $3^3 \times 8$ lattice and the $4^3 \times 16$ lattice, respectively. The $3^3 \times 8$ lattice was first used to observe the tendencies of correlations between squared edge lengths (85) and between 4-volumes (86).

Figure 10 exhibits the correlations $G_{tt}^{(q)}$ of squared edge lengths with edges parallel to the long lattice direction. For the Regge action in the well-defined phase, $\beta < \beta_c \approx 0.1055$, correlations are very short ranged and more or less zero for $d \geq 2$. Arround the critical point $\beta_c$ edges are correlated with a long range and shallow slope. For $\beta > \beta_c$, correlations change considerably and tend to oscillate.

The situation for the action $I_C$ (lower plot of Figure 10) is not very different, but correlations have larger range for the corresponding $\beta$ values and there is no zigzag behavior seen. The oscillating domain for $I_C$ is possibly reached for $\beta > 0.15$, where the average fatness becomes small enough.

In Figure 11 our results for the largest eigenvalue $\gamma_{max}^{(q)}$ of the edge correlation matrix $G_{ab}^{(q)}$ (85) are drawn. Here all 15 edges are involved in contrast to the last plot dealing with edge lengths in the long direction. The maximum eigenvalues indicate two different types of correlations corresponding to the two phases of the system when using $I_R$, whereas for the case of $I_C$ no unequivocal separation into two kinds is possible. Oscillations become more pronounced with increasing gravitational coupling. One should remark that considering the correct geodesic distance could change all pictures.

In constructing volume correlations, the sum over 4-volume elements at site $v$ in (86) was restricted to one hypercube in the direction of $d$. This keeps the observable more localized and leaves the results practically unchanged. Volume correlations $G^{(V)}$ are presented in Figure 12 and behave similar as the eigenvalues $\gamma_{max}^{(q)}$ of the edge length correlations. For the Regge action in the smooth phase they vanish for $d \geq 2$ and oscillate with increasing $\beta$ in the rough phase. The volume correlations for the compact action are nearly proportional to the edge correlation eigenvalues, cf. Figure 11. It is remarkable that the volume correlations for $I_C$ are strictly positive in comparison to those for $I_R$.

Next, in Figures 13 – 15 for the Regge action in the well-defined phase the edge and volume correlations are contrasted with those obtained for a different choice of the cosmological constant, $\lambda = \sigma = 0.25$. Since it was demonstrated in earlier work [30] that one-point expectation values do not change very much in the range $0 \leq \sigma \lesssim 1$, in this way one can directly study the effect of the cosmological constant $\lambda$ on the two-point functions. This subject will be discussed now on the larger lattice.

In general, for correlations (84) one expects the functional form [42]

$$G(d) = C \frac{1}{d^p} e^{-md} \; , \tag{88}$$

which was investigated quantitatively on a $4^3 \times 16$ lattice. Figures 16 – 18 display the results of edge, volume and curvature correlations for the Regge action. In our analysis several fit philosophies have been tried. One can attempt to fit the correlations to a pure exponential function or try a power law. It turned out, that the exponential decay is more prefered suggesting finite effective masses.

For the maximum eigenvalues of $G_{ab}^{(q)}$ the fit parameters are given in Table 2. In both cases, $\lambda = \sigma = 1.0$ and $\lambda = \sigma = 0.1$, a decrease of the effective mass $m^{(q)}$ towards the critical coupling is visible, consistent with a massless particle at criticality. On the other hand, a decrease due to smaller cosmological constant is not observed. The same behavior turns out for the maximum eigenvalues of the deficit angle correlation matrix $G_{ab}^{(\delta)}$.

The volume correlations $G^{(V)}$ show a decreasing mass $m^{(V)}$ as $\lambda$ is reduced, cf. Table 3. When approaching the critical point for constant $\lambda = \sigma = 1.0$ or $\lambda = \sigma = 0.1$, respectively, one finds contrary effects: $m^{(V)}$ increases in the first case for $\beta \to \beta_c \sim 0.108$, but decreases for $\beta \to \beta_c \sim 0.14$ in the latter. Thus, infering a massless scalar particle as expected from weak-field calculations is hardly possible.

The curvature correlations $G^{(R)}$ in Figure 18 behave rather different compared to $G^{(V)}$ or $\gamma_{max}^{(q)}$. At $d = 1$ all values of $G^{(R)}$ are considerably below zero, which means that the local curvatures at neighboring vertices are strongly anti-correlated. Anyhow, for the fit procedure only index distances $d \geq 2$ are taken into account, because correlations at large distances should signal the occurence of massless gravitons. In the case $\lambda = \sigma = 0.1$, $m^{(R)}$ decreases similar as $m^{(V)}$ and $m^{(q)}$ when approaching the critical point. For $\lambda = \sigma = 1.0$, $G^{(R)}$ has strong fluctuations and it was impossible to fit correlations for $\beta = 0.108$. To get an idea of the local curvatures to be correlated in this case, see Figure 19, which at least gives some insight into the anticorrelated nature of the observable. It seems to be too daring to deduce the existence of massless spin-2 exchange particles from our data.

| $\lambda = \sigma$ | 1.0 | 1.0 | 0.1 | 0.1 |
|---|---|---|---|---|
| $\beta$ | 0.1 | 0.108 | 0.1 | 0.14 |
| $C^{(q)}$ | 0.2 | 0.2 | 3.0 | 3.0 |
| $m^{(q)}$ | 0.7 | 0.4 | 2.9 | 2.5 |
| $C^{(\delta)}$ | 0.1 | 0.1 | 0.3 | 0.3 |
| $m^{(\delta)}$ | 1.2 | 0.8 | 2.4 | 2.4 |

Table 2: Fit parameters for the maximum eigenvalue of the edge length and deficit angle correlation matrices to a pure exponential.

| $\lambda = \sigma$ | 1.0 | 1.0 | 0.1 | 0.1 |
|---|---|---|---|---|
| $\beta$ | 0.1 | 0.108 | 0.1 | 0.14 |
| $m^{(V)}$ | 5.7 | 6.1 | 3.4 | 3.1 |
| $m^{(R)}$ | 0.4 | $\times$ | 1.2 | 0.5 |

Table 3: Fit parameters for the volume and curvature correlations to a pure exponential ($C \approx 1$).

To conclude, we have obtained first qualitative results for two-point functions of 4-dimensional simplicial quantum gravity [43]. Geodesic distances are not taken into account, which would be a considerable effort without changing things drastically in the well-defined phase. Although no convincing evidence for the particle spectrum predicted from weak-field approximations could be found, at least simulations with small cosmological constant and almost scale-invariant measure ($\lambda = \sigma = 0.1$) show a decrease of effective masses approaching the critical coupling. For computations applying the uniform measure and the corresponding larger cosmological constant ($\lambda = \sigma = 1.0$) statistics is not sufficient to yield clear signals for curvature correlations. The question of the influence of the cosmological constant on the graviton mass thus remains unanswered.

Recently, interesting results were also reported for volume and curvature correlations with an additional $R^2$ term in the action [42]. There, an approximation for the geodesic distance was applied by measuring the relaxation length of a scalar field. These results are in favor of a massless graviton and thus qualitatively different from our data presented in this Section. However, a direct comparison is not possible at this line because our notions of distances differ.

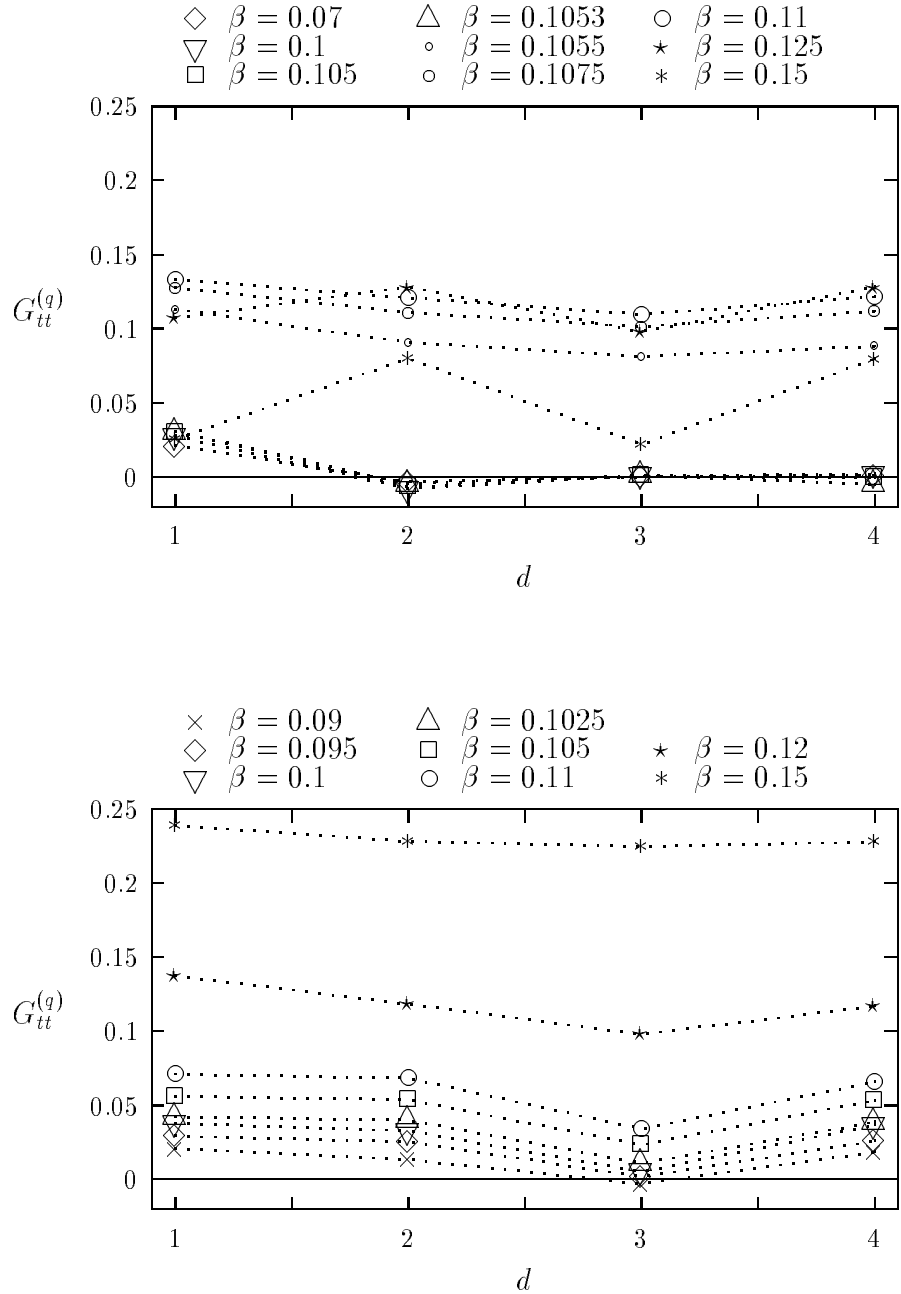

Figure 10: Correlations between edge lengths in the long lattice direction, $G_{tt}^{(q)}$, at different vertices separated by the index distance $d$ for the Regge action $I_R$ (upper picture) and for the compact action $I_C$ (lower picture) with $\sigma = \lambda = 1.0$ on the $3^3 \times 8$ lattice.

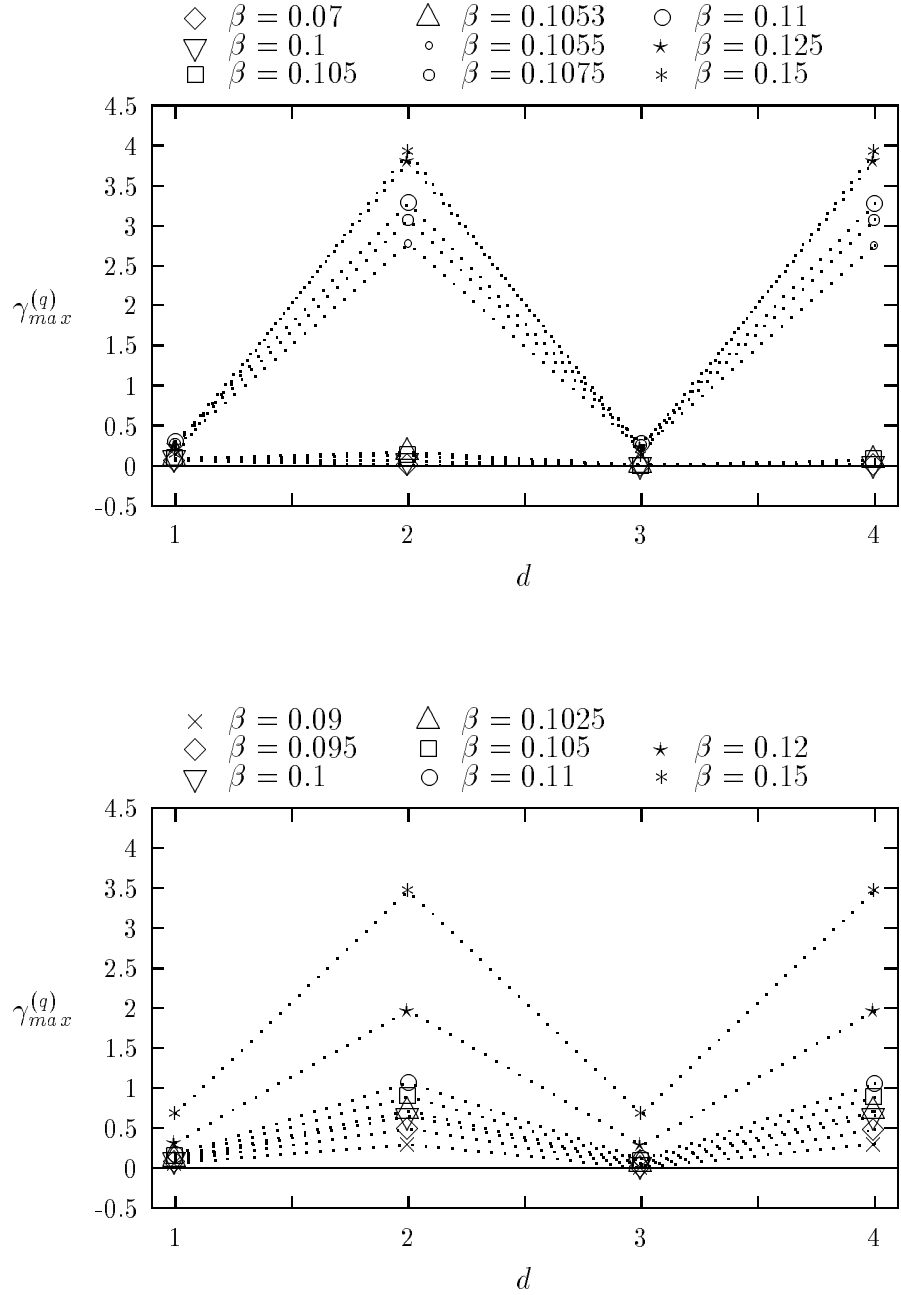

Figure 11: Maximum eigenvalue $\gamma_{max}^{(q)}$ of the edge length correlation matrix $G_{ab}^{(q)}$ for $I_R$ (upper picture) and $I_C$ (lower picture) with $\sigma = \lambda = 1.0$ on the $3^3 \times 8$ lattice.

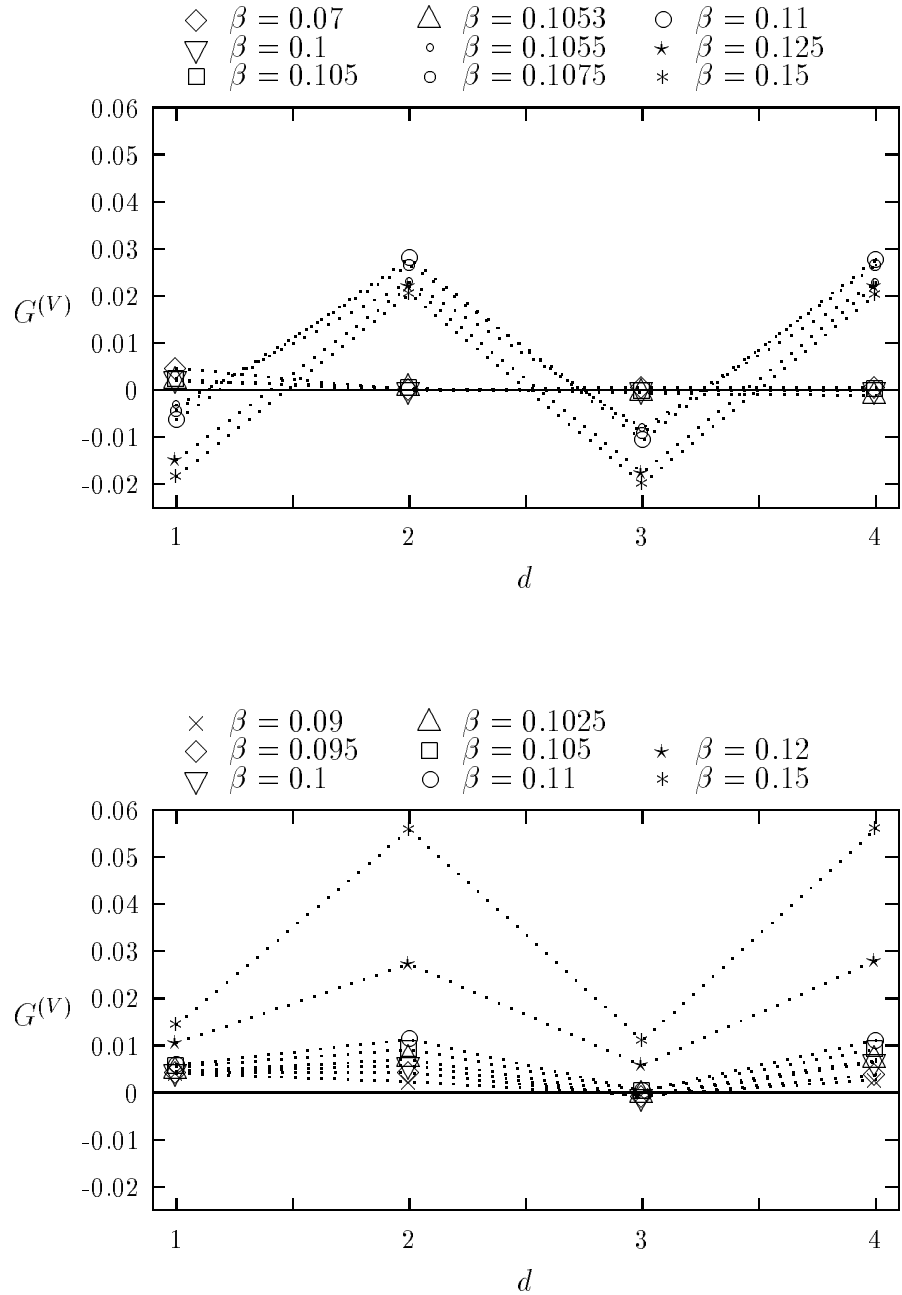

Figure 12: Correlations between 4-volume elements for $I_R$ (upper plot) and $I_C$ (lower plot) with $\sigma = \lambda = 1.0$ on the $3^3 \times 8$ lattice.

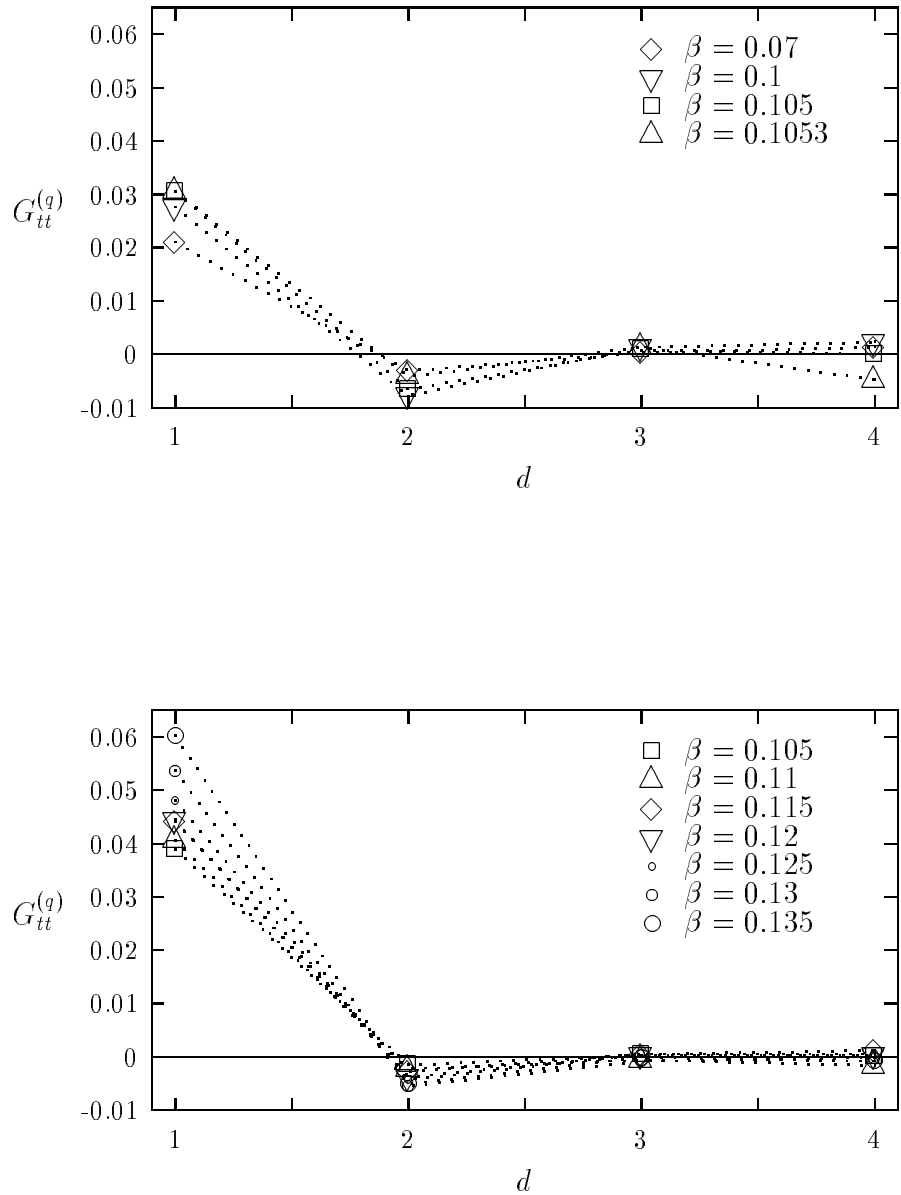

Figure 13: Correlation functions between edge lengths in the long lattice direction at different vertices separated by the index distance $d$. The gravitational coupling lies in the well-defined phase using the Regge action with $\lambda = \sigma = 1.0$ (upper plot) and $\lambda = \sigma = 0.25$ (lower plot).

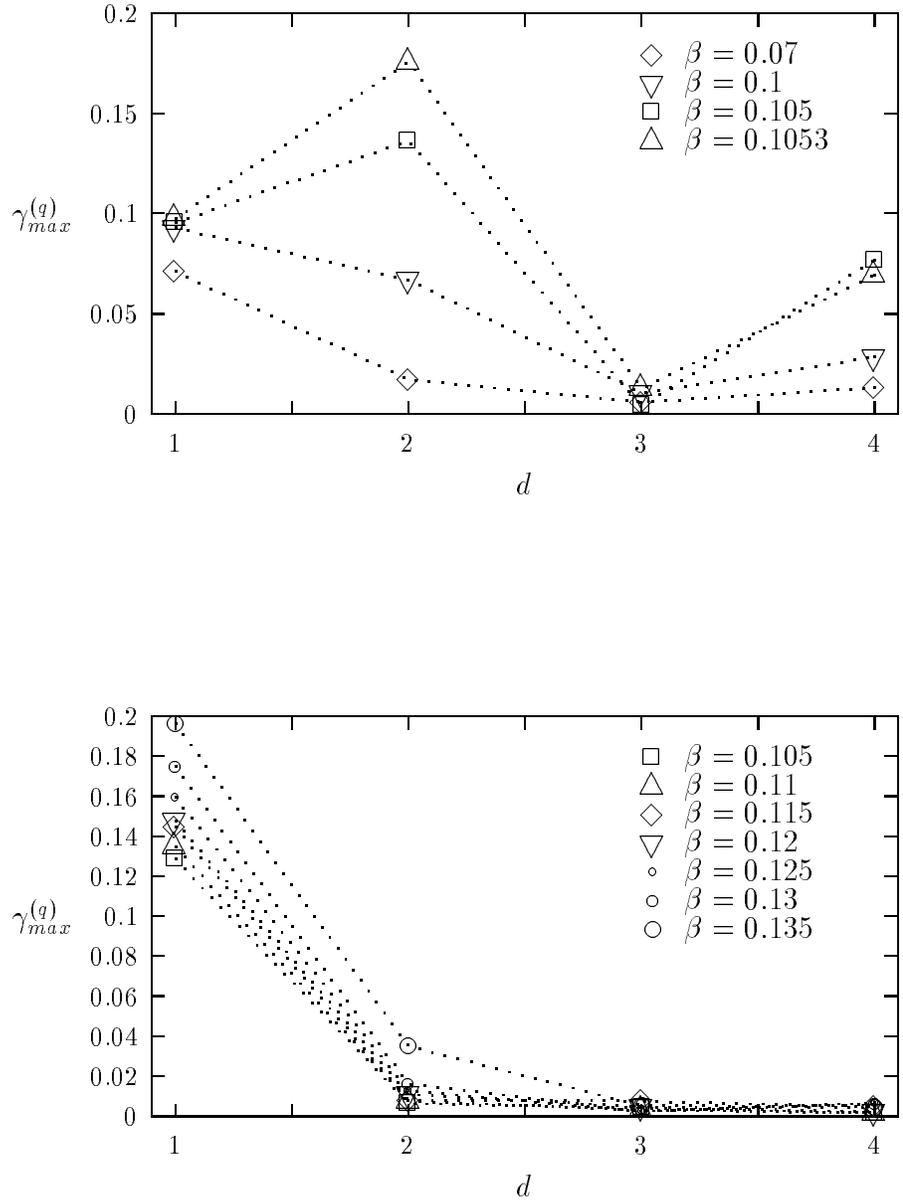

Figure 14: Maximum eigenvalue of $G_{ab}^{(q)}$ versus the index distance $d$ using the Regge action with $\lambda = \sigma = 1.0$ (upper plot) and $\lambda = \sigma = 0.25$ (lower plot).

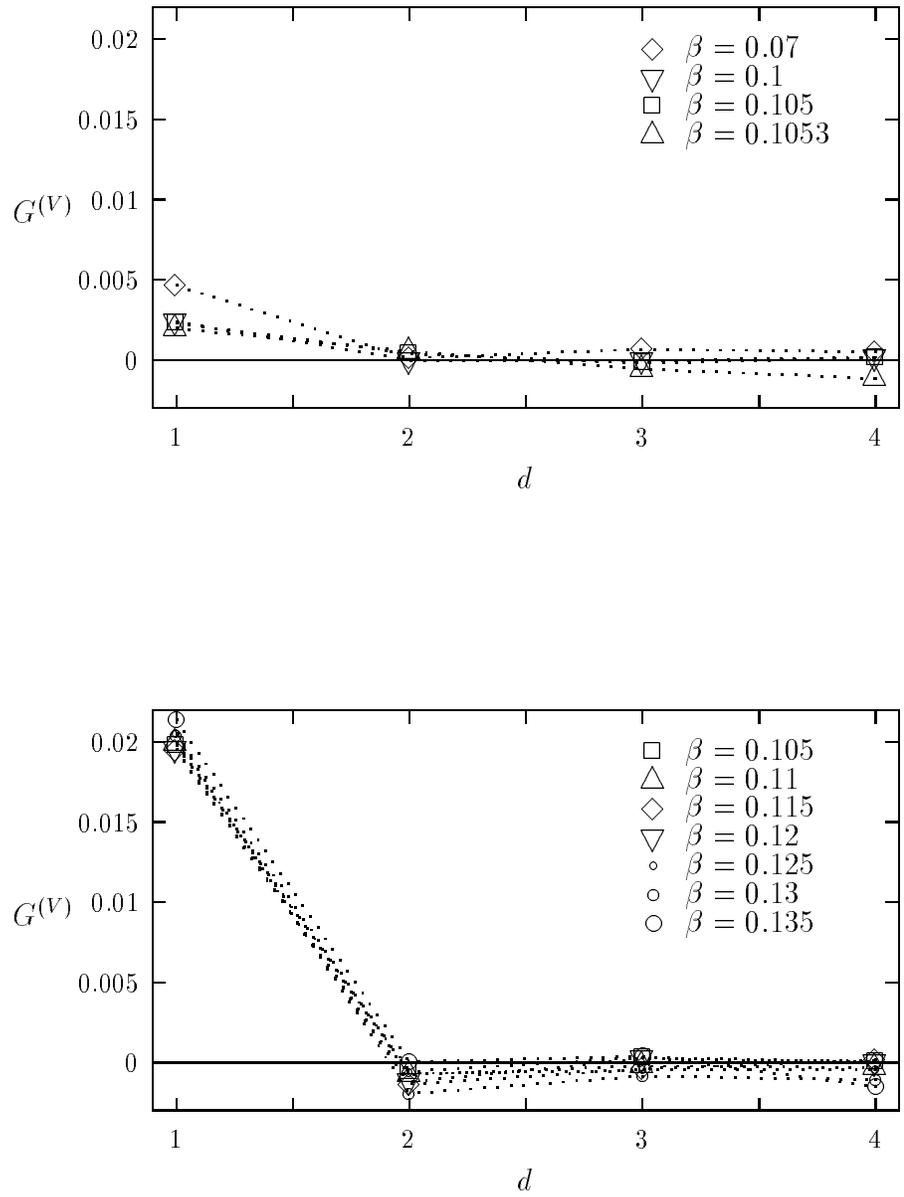

Figure 15: Correlations between 4-volume elements using the Regge action with $\lambda = \sigma = 1.0$ (upper plot) and $\lambda = \sigma = 0.25$ (lower plot).

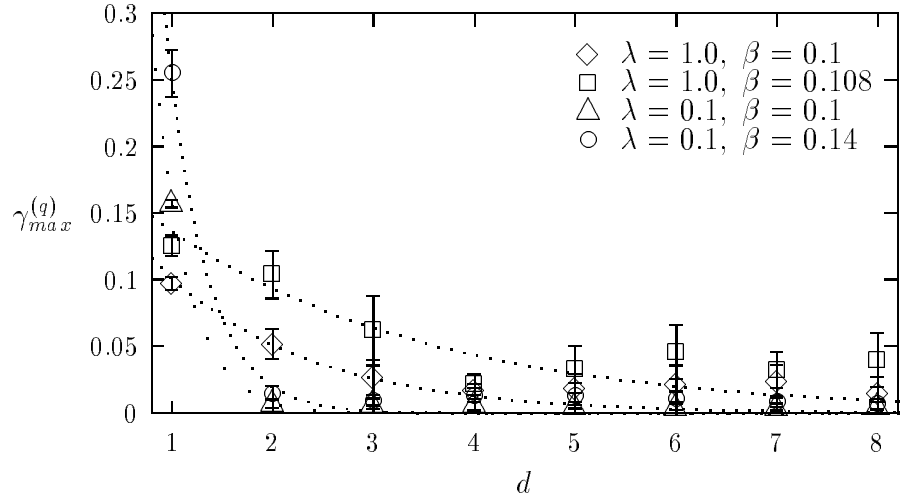

Figure 16: Largest eigenvalue of the edge length correlation matrix versus $d$ for a 4-torus with $4^3 \times 16$ vertices using the Regge action. Two different cosmological constants $\lambda = 0.1, 1.0$ are shown together with one coupling in the well-defined phase, $\beta = 0.1$, and with one coupling near the critical value, $\beta = 0.14, 0.108$.

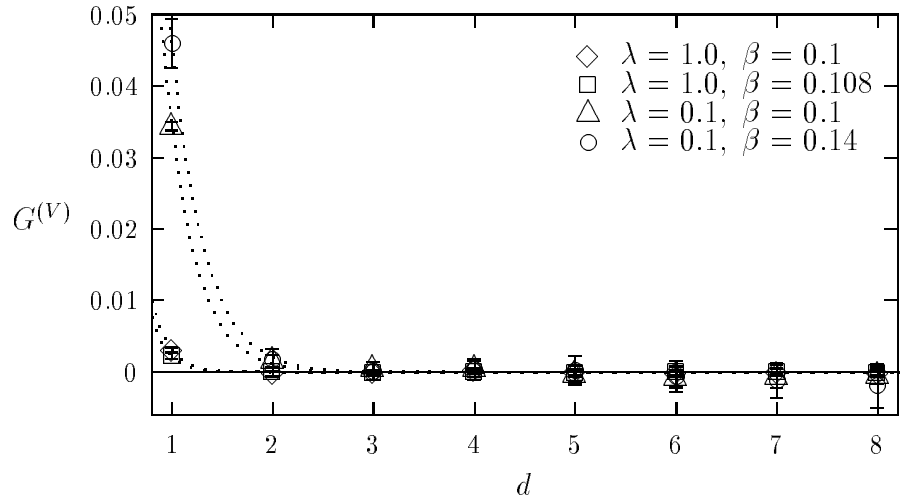

Figure 17: Volume correlations on a 4-torus with $4^3 \times 16$ vertices. Cosmological constants $\lambda = 0.1, 1.0$ are combined with $\beta = 0.1$ and with $\beta = 0.14, 0.108$.

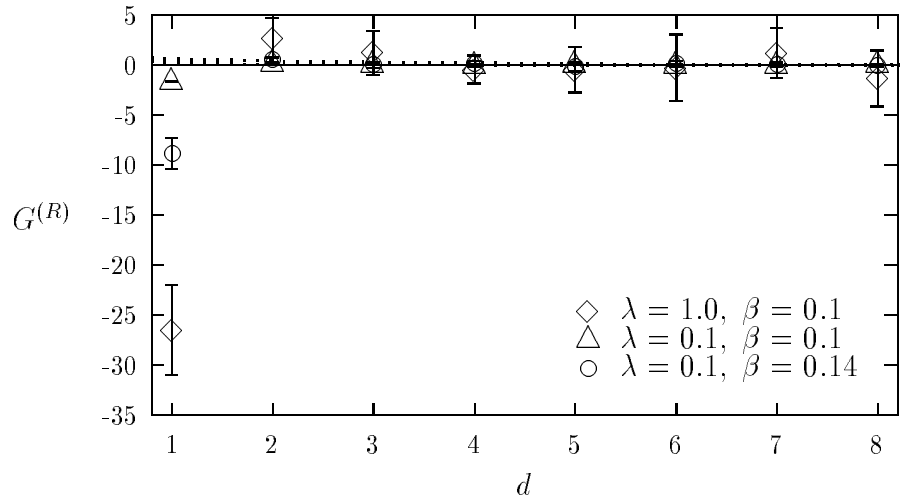

Figure 18: Curvature correlations on a 4-torus with $4^3 \times 16$ lattice. Cosmological constants $\lambda = 0.1, 1.0$ are combined with different gravitational couplings.

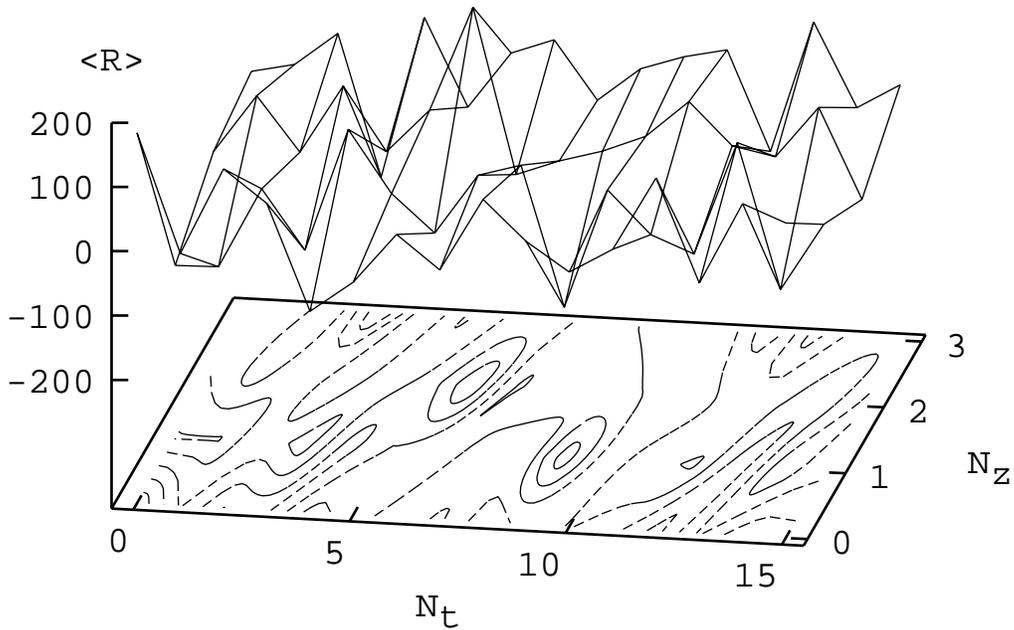

Figure 19: Local curvature of a typical configuration near the critical coupling for $\beta = 0.108$ and $\lambda = \sigma = 1.0$ in the "$(N_z, N_t)$ plane".

# 6 Lattice Gravity as a Spin System

It has been stated that the phase structure of simplicial quantum gravity shows certain similarities to spin glass models [29]. In this spirit, the path integral may be transformed to a partition function of a spin system using certain restrictions on the edges of the lattice [44]. This approach is structurally and computationally much simpler than the Regge calculus or the dynamical triangulation method. It is regarded as a *third lattice gravity approach* and is referred to as *Ising-link quantum gravity* in the literature [45].

## 6.1 Two dimensions

The model is easy to define. All the squared edge lengths are allowed to take only two values
$$q_l = 1 + \epsilon \sigma_l \,, \quad 0 \leq \epsilon < 0.6, \quad \sigma_l \in Z_2 \,, \tag{89}$$
similar to the Regge-Ponzano model [46]. The real parameter $\epsilon$ is restricted to fulfill the Euclidean triangle inequalities for the $q_l$'s so that all $2^{N_1}$ configurations are allowed ($N_1$ is the total number of edges). This is quite different from conventional Regge calculus where many potential updates either violate the triangle inequalities or the manifold property. Furthermore, it may provide a natural measure giving all $2^{N_1}$ configurations equal weight. To use $q_l = k(1 + \epsilon \sigma_l)$ is no more general, as $k$ can be absorbed into a redefinition of $\lambda$.

According to the Gauss-Bonnet theorem the Einstein action in two dimensions is a topological invariant equal to $4\pi$ times the Euler characteristic of the surface. Therefore, for a manifold with fixed topology only the cosmological term remains. To rewrite the action
$$I = \lambda \sum_t A_t \tag{90}$$
in terms of $\sigma_l$ we consider a single triangle (see Figure 20). Its (squared) area

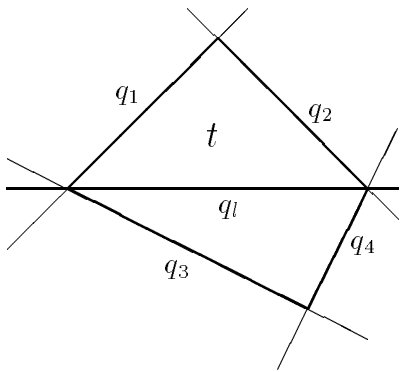

Figure 20: Notation for our triangular lattice. Triangle $t$ consists of edges with the quadratic lengths $q_1$, $q_2$, $q_l$.

can be expressed as

$$A_t^2 = \begin{vmatrix} q_1 & \frac{1}{2}(q_1 + q_2 - q_l) \\ \frac{1}{2}(q_1 + q_2 - q_l) & q_2 \end{vmatrix}$$
$$= \frac{3}{4} + \frac{1}{2}(\sigma_1 + \sigma_2 + \sigma_l)\epsilon + \frac{1}{2}(\sigma_1\sigma_2 + \sigma_1\sigma_l + \sigma_2\sigma_l - \frac{3}{2})\epsilon^2 \,. \quad (91)$$

Expanding $\sqrt{A_t^2}$ the series consists only of terms up to $\sigma^3$ since $\sigma_i^2 = 1$. This suggests the following ansatz

$$A_t = c_0(\epsilon) + c_1(\epsilon)(\sigma_1 + \sigma_2 + \sigma_l) + c_2(\epsilon)(\sigma_1\sigma_2 + \sigma_1\sigma_l + \sigma_2\sigma_l) + c_3(\epsilon)\sigma_1\sigma_2\sigma_l \,. \quad (92)$$

There are only four possible values for the area of a triangle. Computing these areas and comparing with (92) results in the following equations for the coefficients $c_i$

$$\begin{aligned} c_0 &= \frac{1}{16}[2\sqrt{3} + 3f(\epsilon) + 3g(\epsilon)] \\ c_1 &= \frac{1}{16}[2\sqrt{3}\epsilon + f(\epsilon) - g(\epsilon)] \\ c_2 &= \frac{1}{16}[2\sqrt{3} - f(\epsilon) - g(\epsilon)] \\ c_3 &= \frac{1}{16}[2\sqrt{3}\epsilon - 3f(\epsilon) + 3g(\epsilon)] \,, \end{aligned} \quad (93)$$

where $f(\epsilon) = \sqrt{(1-\epsilon)(3+5\epsilon)}$ and $g(\epsilon) = \sqrt{(1+\epsilon)(3-5\epsilon)}$, cf. Figure 21. Hence, one must have $\epsilon < \frac{3}{5}$ for the triangle areas to be real and positive definite. Using

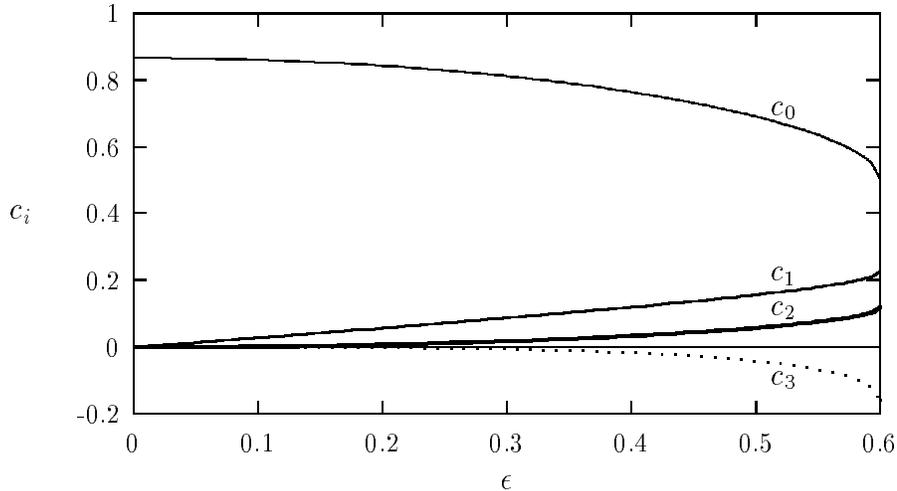

Figure 21: Coefficients $c_i$ ($i = 0, \ldots, 3$) obtained by rewriting the triangle area in terms of $\sigma_l = \pm 1$.

(89) the measure (44) can be replaced by

$$\prod_l \int \frac{dq_l}{q_l^m} \to \sum_{\sigma_l=\pm 1} \exp[-m \sum_l \ln(1+\epsilon\sigma_l)] , \quad (94)$$

and the exponential in terms of $\epsilon$ is reduced to

$$m \sum_l \ln(1+\epsilon\sigma_l) = N_1 m_0(\epsilon) + \sum_l m_1(\epsilon)\sigma_l , \quad (95)$$

with $m_0 = -\frac{1}{2}m\epsilon^2 + O(\epsilon^4)$ and $m_1 = m(\epsilon + \frac{1}{3}\epsilon^3) + O(\epsilon^5)$.
Inserting (92) and (94) into the partition function yields

$$\begin{aligned}
Z &= \sum_{\sigma_l=\pm 1} J \exp\{-\sum_l (2\lambda c_1 + m_1)\sigma_l - \\
&\quad -\lambda \sum_t [c_2(\sigma_1\sigma_2 + \sigma_1\sigma_l + \sigma_2\sigma_l) + c_3\sigma_1\sigma_2\sigma_l]\} \\
&= \sum_{\sigma_l=\pm 1} J \exp\{-\sum_l [(2\lambda c_1 + m_1)\sigma_l + \\
&\quad +\frac{1}{2}\lambda c_2(\sigma_1 + \sigma_2 + \sigma_3 + \sigma_4)\sigma_l + \frac{1}{3}\lambda c_3(\sigma_1\sigma_2 + \sigma_3\sigma_4)\sigma_l]\} , \quad (96)
\end{aligned}$$

with $J = \exp(-\lambda N_2 c_0 - N_1 m_0)$ and $N_2$ the total number of triangles. Thus, the path integral becomes the partition function of a system consisting of a spin $\sigma_l$ at each edge $l$, with an external *magnetic field* and with 2- and 3-spin nearest neighbor interactions. Assigning the spin to the corresponding edge of the originally triangular lattice and drawing the interactions as lines a Kagomé lattice is obtained (see Figure 22).

Removing the term linear in $\sigma$ by a convenient choice of the measure ($m_1 = -2\lambda c_1$) and neglecting 3-spin couplings we get the partition function of an Ising model on a Kagomé lattice. To compute the critical coupling we define $Q = -\frac{1}{2}c_2\lambda$ and transform the partition function $Z_K(Q)$ of the Kagomé lattice to the partition function $Z_D(L)$ of the decorated honeycomb lattice by applying the star-triangle transformation [47]:

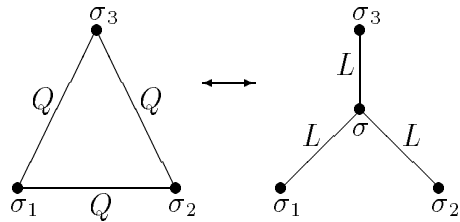

$$\Delta \exp[Q(\sigma_1\sigma_2 + \sigma_2\sigma_3 + \sigma_3\sigma_1)] = \sum_{\sigma=\pm 1} \exp[L\sigma(\sigma_1 + \sigma_2 + \sigma_3)] . \quad (97)$$

For $\sigma_1 = \sigma_2 = \sigma_3 = 1$ and $\sigma_1 = \sigma_2 = -\sigma_3 = 1$ we get $2\cosh(3L) = \Delta\exp(3Q)$ and $2\cosh L = \Delta\exp(-Q)$, respectively. Any other choice of $\sigma_i$ coincides with one of these two cases. Hence, $\Delta$ and $Q$ are determined from

$$\begin{aligned}
\Delta^4 &= e^{4Q}(e^{4Q} + 3)^2 \\
e^{4Q} &= 2\cosh(2L) - 1 . \quad (98)
\end{aligned}$$

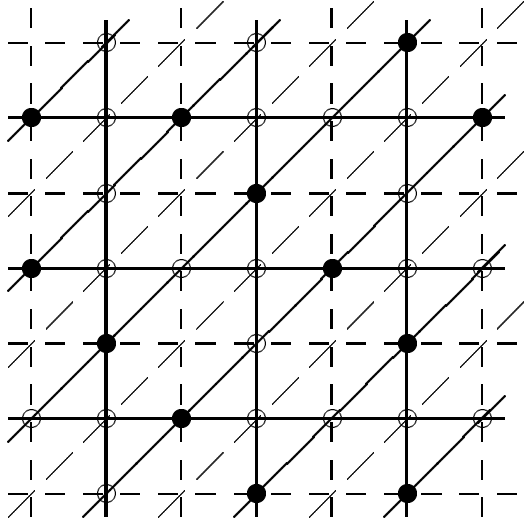

Figure 22: The triangular lattice is drawn by dashed lines and the Kagomé lattice by solid lines. The spins are represented by o and • characterizing a typical configuration discussed in the text.

Carrying out this transformation at every vertex gives $Z_D(L) = \Delta^{\frac{2}{3}N_H} Z_K(Q)$ where $N_H$ is the number of vertices of the honeycomb lattice. Furthermore, we can express $Z_D$ by the partition function $Z_H$ of the conventional honeycomb lattice via the decoration-iteration transformation [47]:

$$\sum_{\sigma=\pm 1} \exp[L\sigma(\sigma_1 + \sigma_2)] = I \exp(K\sigma_1\sigma_2) \ . \tag{99}$$

Similarly, we have

$$I = 2e^K$$
$$e^{2K} = \cosh(2L) \ , \tag{100}$$

and inserting (98) and (100) we obtain

$$Z_D(L) = [e^{2Q}(e^{4Q} + 3)]^{\frac{1}{3}N_H} Z_K(Q) = (2e^K)^{N_H} Z_H(K)$$
$$e^{4Q} = 2e^{2K} - 1 \ . \tag{101}$$

The critical coupling of the honeycomb lattice is given by $e^{2K_c} = 2 + \sqrt{3}$ and thus for the Kagomé lattice

$$e^{4Q_c} = 3 + 2\sqrt{3}$$
$$Q_c = -\tfrac{1}{2}(\lambda c_2)_c = 0.4643 \ . \tag{102}$$

In the ferromagnetic regime $Q > 0$ ($\lambda < 0$) the system shows a $2^{nd}$-order phase transition, whereas the antiferromagnetic regime $Q < 0$ ($\lambda > 0$) is governed by frustration even at zero temperature.

Switching on the symmetry-breaking 3-spin interaction we performed numerical simulations of the system with toroidal topology in the coupling plane $(Q_2 = -\frac{1}{2}\lambda c_2, Q_3 = -\frac{1}{3}\lambda c_3)$. Figure 23 shows the spin expectation value in a certain range of coupling constants. For $Q_3 = 0$ the conventional Ising model on a Kagomé lattice is recovered and for $Q_3 \neq 0$, $Q_2 < 0$ the 3-spin coupling removes the frustration leading to ordered phases. Depending on sign($Q_3$) the spins denoted in Figure 22 by open dots take the values $\sigma = \pm 1$ whereas the full dots take $\sigma = \mp 1$ giving rise to an expectation value $\langle \sigma \rangle = \pm \frac{1}{3}$.

Now return to the application of the spin system to the gravitational system. To investigate different types of measures the external field term has to be taken into account. Further the couplings depend on the value of $\epsilon$ according to (93). Figure 24 depicts $\langle \sigma \rangle$ versus $\epsilon$ for the uniform and the scale invariant measure and for the measure leading to a cancellation of the term linear in $\sigma$. The upper curve can be read off directly from Figure 23 and represents only a small detail near the origin. The 3-spin coupling is in the entire range of $\epsilon$ not strong enough to remove the frustration and therefore $\langle \sigma \rangle \approx 0$. Both other curves show a favored occurence of negative spins, corresponding via $q_l = 1 + \epsilon \sigma_l$ to short edge

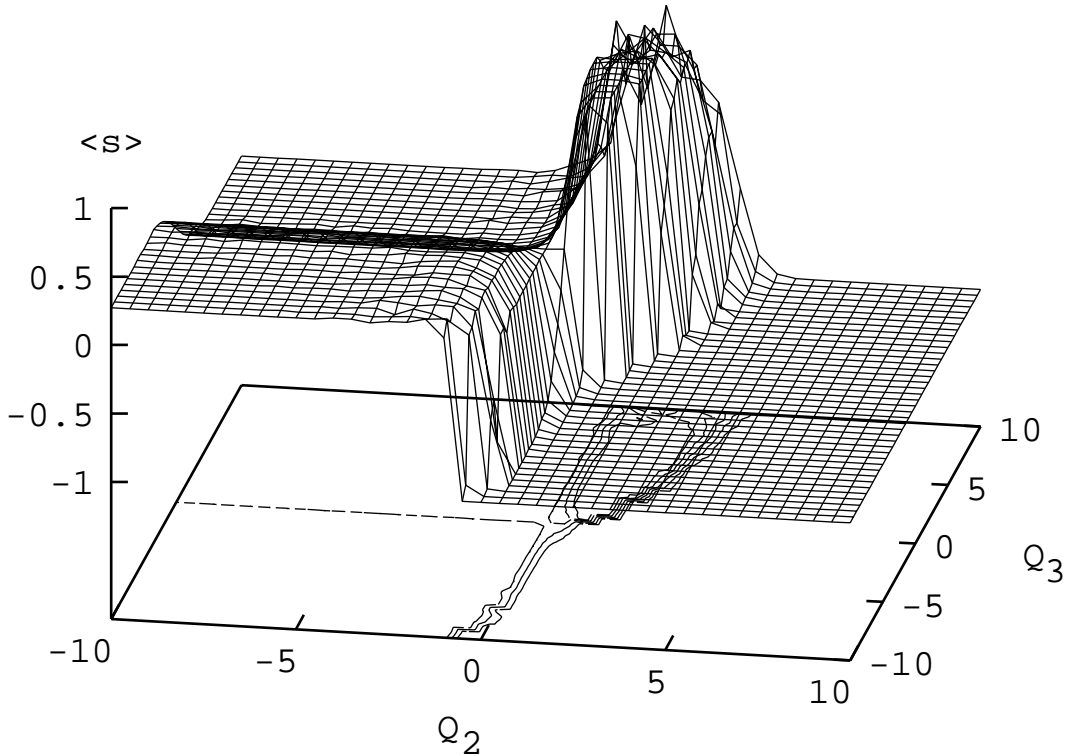

Figure 23: Spin expectation value as a function of the 2- and 3-spin couplings $Q_2$ and $Q_3$, respectively.

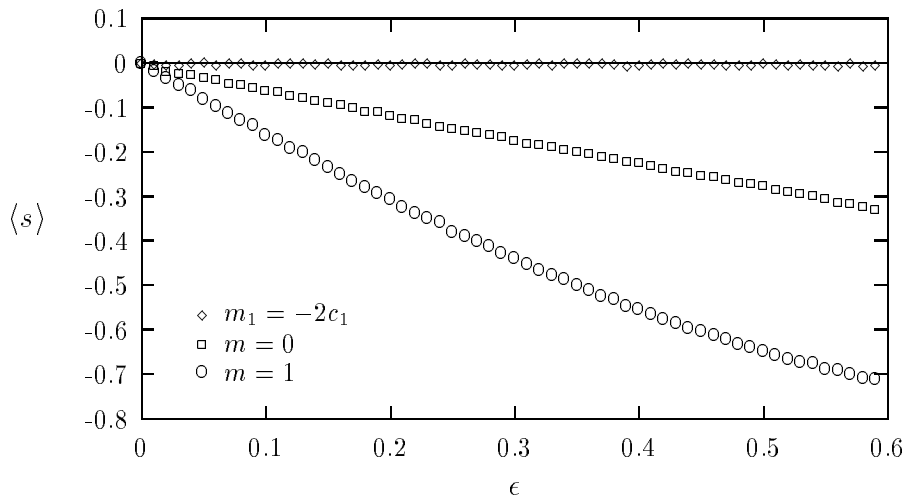

Figure 24: Spin expectation value versus $\epsilon$ for $m_1 = -2c_1$ (no external field), $m = 0$ (uniform measure) and $m = 1$ (scale invariant measure). In all cases the cosmological constant is set to $\lambda = 1$.

lengths reflecting the expected tendency of the lattice to shrink [48].

An extension of our approach to more than two edge lengths can be easily performed and it is straightforward to include additional scalar fields or physical spin fields. A generalization of the trivial 2-dimensional Regge-Einstein action to higher dimensions is of course possible, but becomes very complicated. Notice that in two dimensions only the three edges of a triangle are coupled in the action. In three dimensions one is faced with terms of $6^{th}$ order at least, and with additional contributions from the Regge action. Recently, the Ising-link model in three dimensions has been analyzed via mean field techniques as well as with the Monte Carlo method [45]. However, critical exponents and the behavior of the curvature at the transition point differ from that found for Regge theory with continuously varying edge lengths.

## 6.2 Four dimensions

The situation is even more complicated in four dimensions where one has to deal with 10 edges in each simplex. Nevertheless, numerical simulations can be made very efficient by implementing look-up tables and using the heat bath algorithm. A spin update consists of choosing a particular spin, calculating the change in the action if the spin takes on its other possible value, and accepting the new spin value with probability to the exponential of the negative change in the action.

In the actual computations we replaced (89) by $q_l = b_l(1 + \epsilon \sigma_l)$ because a 4-dimensional Regge skeleton with equilateral simplices cannot be embedded in flat space. Therefore, according to the indexing scheme of Section 3.1, $b_l$ takes different values depending on the type of the edge $l$. In particular $b_l = 1, 2, 3, 4$ for edges, face diagonals, body diagonals, and the hyperbody diagonal of a hypercube, respectively. First preliminary studies show an interesting phase structure. The range of the gravitational coupling $\beta$ is extended also to negative values, because there seems to be no particular reason to stay in the Euclidean sector.

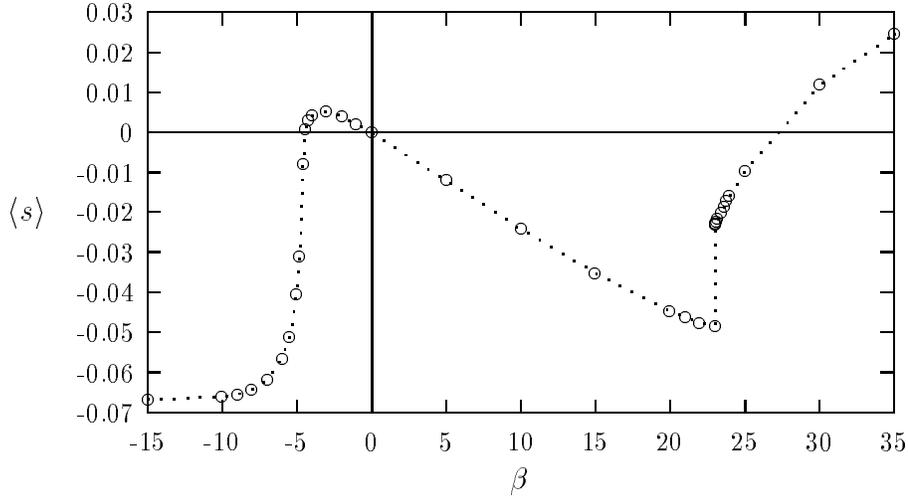

Figure 25: Spin expectation value as a function of the gravitational coupling parameter for Ising-link quantum gravity in four dimensions.

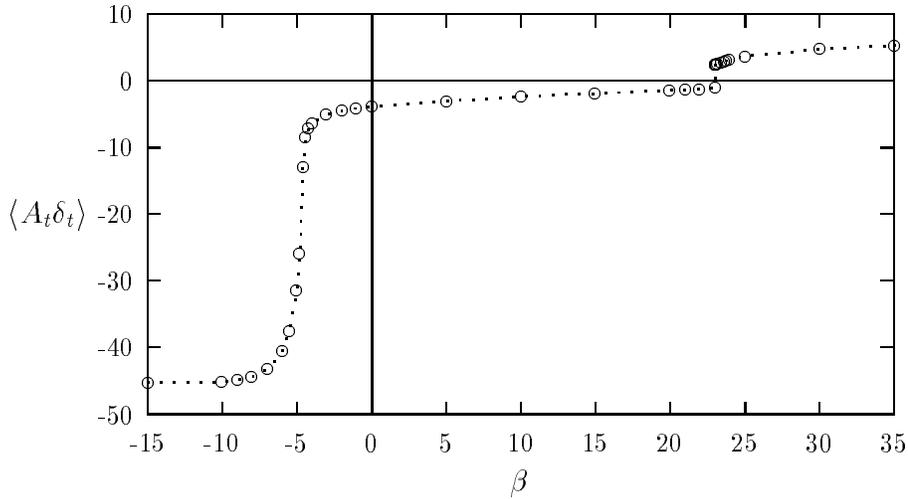

Figure 26: Average action for simulations with different couplings $\beta$ and $\epsilon = 0.0875$.

The spin expectation value of a 4-dimensional toroidal lattice with $8^4$ vertices is depicted in Figure 25. The parameter $\epsilon$ was set to 0.0875, $\lambda = 0$ and $\sigma = 1$. The system seems to undergo phase transitions at two different points, first in the negative region at $\beta \approx -4.7$ and second at a large positive coupling $\beta \approx 23$.

In Figure 26 the expectation value of the action versus the gravitational coupling $\beta$ is presented. The resemblance of the curve for $\beta > 0$ with those obtained by continuously varying edge lengths is very encouraging. One has to determine the value of the critical point and the order of the phase transition in the limit $N_0 \to \infty$ in more detail. If both theories belong to the same universality class, one could perform calculations much easier and faster than with conventional Regge calculus. But to abandon dwelling on speculations, this is a question to be answered in future work.

# 7   Conclusion

The Regge calculus provides a direct route to systematic approximations of the quantum-gravity path-integral. Unfortunately, there are almost no analytical results or experimental data that could guide us and thus, numerical simulations have to be interpreted very cautiously. In this spirit let us try to summarize.

Although the Regge-Einstein action $I_R$ and the "compact" action $I_C$ are unbounded the entropy of the system stabilizes expectation values leading to the existence of a well-defined phase with small average curvature. Increasing the gravitational coupling $\beta$ the system undergoes a phase transition, which is abrupt for $I_R$ and smoother for $I_C$. This might be an important result for the continuum limit of simplicial quantum gravity.

To get an idea of the details of the interaction, correlation functions between several observables like edge lengths, local volume elements, and local curvatures have been computed. Most correlation functions exhibit a decrease of the corresponding effective mass towards the critical coupling. The influence of the cosmological constant shows no universal trend for the masses. However, no clear evidence for long range correlations is found, which should have indicated the occurence of massless particles. At present it is too early to preclude the existence of a massless graviton because with the appropriate action, measure, the true geodesic distance, and better statistics the reliability of the two-point functions will improve.

Finally, a new method to investigate the phase structure of simplicial quantum gravity was introduced. Restricting the squared edge lengths of the lattice to take only two values the path integral can be transformed to the partition function of a spin system. This allows to handle the system in a very economic way from an analytical as well as from a numerical point of view. Especially, if both quantum gravity and the corresponding spin system lie in the same universality class, this seems to be very promising for future investigations.

If no $2^{nd}$ order phase transition will be found, an intriguing idea would be to consider the lattice spacing as a fundamental length. Then one does not need to send the lattice spacing to zero ($q \to 0$) as in usual lattice gauge theory, but there exists a limit $q_0 = \text{finite} = \mathcal{O}(L_P)$. A further appealing idea is the construction of a grand unified theory of the fundamental forces by introducing additional matter fields, e.g. $SU(2)$ gauge fields [49]. Then one has to try to tune the hadronic masses such that they are sufficiently small compared to the Planck mass. Also, the curvature of the lattice has to approach zero and the geometry to become flat on the hadronic length scale. The influence of matter fields on the graviton mass would be of tremendous interest.

**Acknowledgement**

The friends to whom I am most grateful and without whom this work would have been impossible are Harald Markum, Wolfgang Beirl, Erwin Gerstenmayer, and Balasubramanian Krishnan.

# List of Figures

## List of Tables

# References

[1] C. Isham, in *Recent Aspects of Quantum Fields*, eds. H. Mitter and H. Gausterer, Lect. Notes in Phys. 396 (Springer, Berlin, 1992) 123; Proc. of the $11^{th}$ Int. Conf. on General Relativity and Gravitation, ed. M. MacCallum (Cambridge Univ. Press, Cambridge, 1987) 99; gr-qc/9310031

[2] E. Alvarez, Rev. Mod. Phys. 61 (1989) 561

[3] G. Gibbons and S. Hawking, *Euclidean Quantum Gravity* (World Scientific Press, Singapore, 1993)

[4] M. Goroff and A. Sagnotti, Phys. Lett. 160B (1985) 81; Nucl. Phys. B226 (1986) 709

[5] P. West, *Introduction to Supersymmetry and Supergravity* (World Scientific Press, Singapore, 1986)

[6] B. DeWitt, Phys. Rev. 160 (1967) 1113

[7] A. Ashtekar, Phys. Rev. Lett. 57 (1986) 2244

[8] C. Rovelli and L. Smolin, Phys. Rev. Lett. 61 (1988) 1155

[9] J. Hartle and S. Hawking, Phys. Rev. D28 (1983) 2960

[10] T. Regge, Nuovo Cimento 19 (1961) 558

[11] J. Hartle, J. Math. Phys. 26 (1985) 804; 27 (1986) 287; 30 (1989) 452

[12] G. Feinberg, R. Friedberg, T.D. Lee, and H.C. Ren, Nucl. Phys. B245 (1984) 343

[13] K. Fredenhagen and H. Haag, Comm. Math. Phys. 108 (1987) 91

[14] M. Green, J. Schwarz, and E. Witten, *Superstring Theory* (Cambridge Univ. Press, Cambridge, 1986)

[15] R. Schoen and S.T. Yau, Phys. Rev. Lett. 42 (1979) 547

[16] E. Fradkin and G. Vilkoviski, Phys. Rev. D8 (1973) 4241

[17] T. Eguchi, P. Gilkey, and A. Hanson, Phys. Rep. 66 (1980) 213

[18] K. Schleich and D. Witt, gr-qc/9307015; gr-qc/9307019

[19] C. Misner, Rev. Mod. Phys. 29 (1957) 497

[20] L. Faddeev and V. Popov, Sov. Phys. Usp. 16 (1974) 777

[21] P. Menotti, Nucl. Phys. B (Proc. Suppl.) 17 (1990) 29

[22] C. Nash and S. Sen, *Topology and Geometry for Physicists* (Academic Press, London, 1983)

[23] N. Christ, R. Feinberg, and T. D. Lee, Nucl. Phys. B202 (1982) 89

[24] M. Roček and R. Williams, Phys. Lett. B104 (1981) 31; Z. Phys. C21 (1984) 371

[25] H. Hamber, in *Critical Phenomena, Random Systems, Gauge Theories*, Proc. of the Les Houches Summer School, eds. K. Osterwalder and R. Stora (North Holland, Amsterdam, 1986) 375;
H. Hamber and R. Williams, Nucl. Phys. B248 (1984) 392; Phys. Lett. 157B (1985) 368; Nucl. Phys. B269 (1986) 712

[26] J. Cheeger, W. Müller, and R. Schrader, Comm. Math. Phys. 92 (1984) 405

[27] M. Caselle, A. D'Adda, and L. Magnea, Phys. Lett. B232 (1989) 457

[28] J. Fröhlich, in *Non-perturbative Quantum Field Theory - Mathematical Aspects and Applications*, Advanced Series in Mathematical Physics, Vol. 15, ed. A. Wightman (World Scientific, Singapore, 1992) 523

[29] B. Berg, in *Particle Physics and Astrophysics*, Proc. of the XXVII. Int. Universitätswochen für Kernphysik, eds. H. Mitter and F. Widder (Springer, Berlin, 1989) 223; Phys. Rev. Lett. 55 (1985) 804; Phys. Lett. B176 (1986) 39

[30] W. Beirl, E. Gerstenmayer, and H. Markum, Nucl. Phys. B (Proc. Suppl.) 26 (1992) 575; Phys. Rev. Lett. 69 (1992) 713; in *Vistas in Astronomy*, Vol. 37 (1993) 605;
W. Beirl, E. Gerstenmayer, H. Markum, and J. Riedler, Nucl. Phys. B (Proc. Suppl.) 30 (1993) 764; Phys. Rev. D, in press (hep-lat/9402002)

[31] F. David, Nucl. Phys. B257 (1985) 45; 543;
J. Ambjørn, B. Durhuus, and J. Fröhlich, Nucl. Phys. B257 (1985) 433;
V. Kazakov, I. Kostov, and A. Migdal, Phys. Lett. 157B (1985) 295;
M. Agishtein and A. Migdal, Int. J. Mod. Phys. C1 (1990) 165;
J. Ambjørn and S. Varsted, Phys. Lett. B266 (1991) 285; Nucl. Phys. B373 (1992) 557

[32] J. Alexander, Ann. Math. 31 (1930) 292

[33] H. Römer and M. Zähringer, Class. Quant. Grav. 3 (1986) 897

[34] A. Nabutovsky and R. Ben-Av, Comm. Math. Phys. 157 (1993) 93

[35] S. Metropolis, A. Rosenbluth, A. Teller, and E. Teller, J. Chem. Phys. 21 (1953) 1087

[36] N. Barth and S. Christensen, Phys. Rev. D28 (1983) 1876

[37] W. Beirl, H. Markum, and J. Riedler, *Simplicial Path Integrals on Nonregular Triangulations*, in preparation

[38] S. Weinberg, *General Relativity An Einstein Centenary Survey*, eds. S. Hawking and W. Israel (Cambridge Univ. Press, Cambridge, 1979) 790

[39] M. Martinelli and A. Marzuoli, Proc. of the $4^{th}$ Marcel Grossmann Meeting on General Relativity, ed. R. Ruffini (Elsevier, Amsterdam, 1986) 1173

[40] M. Veltman, in *Methods in Field Theory*, Proc. of the Les Houches Summer School, eds. R. Balian and J. Zinn-Justin (North Holland, Amsterdam, 1975) 265

[41] P. van Nieuwenhuizen, Nucl. Phys. B60 (1973) 478

[42] H. Hamber, Nucl. Phys. B (Proc. Suppl.) 25A (1992) 150; Nucl. Phys. B400 (1993) 347; hep-th/9311024

[43] W. Beirl, H. Markum, and J. Riedler, Nucl. Phys. B (Proc. Suppl.) 34 (1994) 736

[44] W. Beirl, H. Markum, and J. Riedler, Proc. of the Workshop on Quantum Field Theoretical Aspects of High Energy Physics, eds. B. Geyer and E.-M. Ilgenfritz (Leipzig 1993) 287; Int. J. Mod. Phys. C, in press (hep-lat/9312055)

[45] T. Fleming, M. Gross, and R. Renken, hep-lat/9401002

[46] G. Ponzano and T. Regge, in *Spectroscopic and Group Theoretical Methods in Physics*, ed. F. Bloch (North Holland, Amsterdam, 1968)

[47] R. Baxter, *Exactly Solved Models in Statistical Mechanics* (Academic Press, London, 1982);
I. Syozi, in *Phase Transitions and Critical Phenomena*, Vol. 3, eds. C. Domb and M. Green (Academic Press, London, 1974) 269

[48] H. Hamber and R. Williams, Nucl. Phys. B267 (1986) 482

[49] B. Berg, B. Krishnan, and M. Katoot, Nucl. Phys. B404 (1993) 359;
B. Berg and B. Krishnan, Phys. Lett. B318 (1993) 59